# A novel hybrid neural network of fluid-structure interaction prediction for two cylinders in tandem arrangement

Yanfang Lyu[a,b,c], Yunyang Zhang[b,c], Zhiqiang Gong[b,c], Xiao Kang[d,*], Wen Yao[b,c,*], Yongmao Pei[a]

[a] State Key Laboratory for Turbulence and Complex Systems & Center for Applied Physics and Technology, College of Engineering, Peking University, Beijing 100871, China
[b] Defense Innovation Institute, Chinese Academy of Military Science, Beijing 100071, China
[c] Intelligent Game and Decision Laboratory, Beijing 100071, China
[d] Institute of Advanced Structure Technology, Beijing Institute of Technology, Beijing 100081, China
* Corresponding Author: kangxiaoyx@163.com; wendy0782@126.com

**Nomenclature**

| | |
|---|---|
| $f^w$, $f^F$ | Trainable function relationships of MLP and USFNO-FConvLSTM |
| $t_i$ or $T_i$ | *ith* time step |
| $N$ | Number of time step |
| $F_i$ | Flow field state at $t_i$ time step |
| $W_i$ | Wall shear stress at $t_i$ time step |
| $S_i$, $y_i$, $\dot{y}_i$ | Structure response, displacement and velocity at $t_i$ time step |
| $P$ | Pressure field or surface pressure |
| $WS$ | Wall shear stress |
| $F_b$, $F_{b\text{-}p}$, $F_{b\text{-}wall}$ | Surface fluid force, force integrated by Surface pressure or Wall shear force |
| NN | Neural network |
| $U_F$, $U_{F\text{-}S}$ | Low-dimensional fluid features, feedback of structural oscillation |
| $\mathcal{L}_{\text{stress}}$, $\mathcal{L}_{\text{Field}}$ | L1 loss function of MLP and USFNO-FConvLSTM |
| $v(x)$, $Q(x)$ | Input and output information of USFNO-FConvLSTM |
| $z_{l_0}$ | Initial input information of the Fourier layer |
| $P_1$, $P_2$ | Linear layer |
| $R_1$, $R_2$ | Reshape function |
| $\sigma$ | Activation function |
| $d_c$ | Channel number of images |
| $L$, $M$ | Number of Conv-Fourier layers and DeConv-Fourier layers |
| $z_{l_L}^{\mathcal{C}}$, $z_{m_M}^{\mathcal{D}}$ | Output of Encoder and Decoder |
| $\mathcal{C}$, $\mathcal{D}$ | Decrease and increase in image resolution |
| $z_T$ | FConvLSTM |
| $I$ | Iterative architecture |
| $\mathcal{F}$, $\mathcal{F}^{-1}$ | Fourier transform and its inverse transform |
| $\mathcal{K}$ | Integral operator |
| $k(x, y)$ | Kernel |
| $k_{max}$ | Number of Fourier modes |
| $[X_1, X_2]$ | Two-dimensional discretization image matrix |
| $(n_x \times n_y)$ | Image revolution |
| $k_m(k_1, k_2, k_3, k_4)$ | The variable coefficients of fourth-order Runge-Kutta |

**ABSTRACT**

Deep learning has shown promise in improving computing efficiency while ensuring modeling accuracy in fluid-structure interaction (FSI) analysis. However, its current capabilities are limited when it comes to constructing multi-object coupling systems with dynamic boundaries. To address such limitation, a novel FSI neural solver integrated by a fluid deep learning model with multi-time steps and a structural dynamic solver is proposed to accurately and reliably predict the vortex-induced vibration (VIV) evolution for two cylinders in tandem. This well-designed model in the form of end-to-end can precisely predict the instantaneous flow field state at the subsequent time by coupling the temporal flow fields of historical multi-time sequences and the current structural responses, moreover, derives the structural state at the next time. Furthermore, the novel fluid deep learning model consists of a wall shear model utilizing a multilayer perception network and flow field model with U-shaped architecture jointing the Fourier neural operator and modified convolution long-short term memory model. Both models effectively capture coupling transfer forces and predict instantaneous flow fields, with the latter demonstrating superior accuracy compared to Convolutional Neural Network- or Unet-based models with similar parameters. The prediction speed of the proposed models realizes an improvement of over 1000 times compared with the numerical simulation. Significantly, the proposed FSI neural model demonstrates exceptional capability in constructing the nonlinear complex multi-vibration systems and has substantial potential for advancing FSI modeling of flexible structures featuring pronounced nonlinear deformation boundaries.

**Keywords**: Flow-induced interaction, fluid mechanics, two tandem cylinders, Fourier neural operator, convolutional long-short term memory model.

## 1. Introduction

FSI, as a complex phenomenon of interactive transfer of physical information between flow fluid and structure, has widely existed in aerospace (He et al., 2022), marine (Sun et al., 2022), civil engineering (Pospíšil et al., 2006), and other practical applications (Lv et al., 2021). Owing to the non-negligible significance of FSI, a considerable number of scholars have carried out abundant theoretical research, experimental measurements (Zhao et al., 2014; Khalak & Williamson, 1997; Huang et al., 2021), and numerical simulations (Mendes & Branco, 1999; Kim & Choi, 2019) on its mechanism and application. Among them, field experiments such as wind tunnel testing (Zarruk et al., 2014) and sea trials (Sun et al., 2022) frequently necessitate considerable allocations of human and material resources. To address the costly experimental processes, commonly used numerical simulation methods including Arbitrary Lagrangian-Eulerian (Sarrate et al., 2001), Level-set (He & Qiao, 2011), Immersed boundary method (IBM) (Borazjani et al., 2008), phase-field modeling (Sun et al., 2014), and so on, have been employed as alternatives. Nevertheless, for complex bidirectional FSI, high-precision mesh modeling is indispensable for accurately transmitting instantaneous boundary information between the fluid domain and the structure, increasing computing time and resource usage, especially in Computational Fluid Dynamics (CFD). In comparison to the time-consuming CFD solvers, the derivation of structural response exhibits relatively higher computational efficiency, meaning that it is crucial to accelerate the solving speed for the fluid part. To tackle the above challenges, surrogate-based models applied in fluid dynamics or FSI systems were proposed and then advanced rapidly in recent years (Zhang et al., 2020; Hidayat & Ariwahjoedi, 2010; Wang et al., 2018; Jin et al., 2018). It can improve the operation efficiency while maintaining the calculation accuracy.

As a typical surrogate-based modeling technology, machine learning methods, such as polynomial chaos expansions (Wu et al., 2018), Kriging model (Vazquez et al., 2006), support vector machine (Asefa et al., 2006), and artificial neural networks (Lv et al., 2021), have been successfully applied to reconstruct and predict fluid flow variations that are approximate to experiments or simulations. The above traditional methods can construct appropriate surrogate models utilizing available labeled data, however, have limited representational ability for strong nonlinear fluid dynamics, especially for the intricate FSI systems. Progressively, existing deep neural network (DNN) models featuring multiple layers have great potential to fit high-dimensional strong nonlinear systems for steady flow field (without regard to temporal progression) (Shirvani et al., 2024; Du et al., 2022) and unsteady flow fluid (consider time evolution) (Srinivasan et al., 2019; Carlberg et al., 2019). For steady state, multilayer perceptron (MLP) with multiple layers and convolutional neural networks (CNNs) with extraction of local and global physical field features both present excellent performance in flow field reconstruction and prediction tasks based on sparse measuring points or different environmental conditions (Shirvani et al., 2024; Du et al., 2022; Liu et al., 2021; Hui et al., 2020). Traditional unsteady flow field studies have mostly focused on the laminar/turbulent evolution without/with the stationary rigid body (Beneddine et al., 2017; Ma et al., 2020) and spatial-temporal physical evolvement involving the vibrating rigid object or deformed flexible body (Ji & Huang, 2017; Martini et al., 2021). The latter involves dynamics boundary and belongs to the category of FSI phenomenon. It includes VIV of single/multiple cylinders, flutter airfoils, wiggling hydrofoils, and so on in the marine and aerospace fields (Zhao et al., 2014; Zarruk et al., 2014). Recently, time sequence neural networks, such as Recurrent Neural Network (RNN) networks (Chaki et al., 2020) and Long Short-Term Memory (LSTM) networks (Cheng et al., 2021), have been presented and proved to be applicable for modeling unsteady flow field prediction in future occasions. LSTM has excellent capability in capturing the temporal evolution characteristics of flow field and force between successive time steps within nonlinear systems, effectively bridging historical and forecasted dynamics (Han et al., 2021; Kou & Zhang, 2021). Furthermore, several studies have utilized Transformer (Ye et al., 2022) and Graph Neural Network (GNN) (Gao & Jaiman, 2022) to construct the spatial-temporal prediction.

For the modeling of unsteady flow fluid or FSI systems, existing deep time series models are mainly divided into two categories: (1) Containing merely the unsteady flow fields as the training data. (2) Determining the dynamic or displacement of structure as the modeling data.

For the first category where models are designed to forecast the instantaneous flow field state and its evolution over time, the focus is on capturing the dynamic behavior of fluids around stationary bluff bodies or within turbulent processes. On the basis of RNN, Sandeep et al. (2021) presented two hybrid data-driven models respectively realized via proper orthogonal decomposition (POD) and convolution recurrent autoencoder network (CRAN) to predict pressure and velocity fields around isolated and double cylinder configurations. Eivazi et al. (2020) constructed an Autoencoder-LSTM network to emulate velocity field evolution around a stationary cylinder and the oscillating airfoil, and to predict the next time step through a sequence of existing flow features. Analogously, Yousif & Lim (2021) integrated POD with LSTM to forecast future temporal coefficients, subsequently reconstructing the instantaneous turbulent flow field around a wall-mounted square cylinder at Reynolds number $Re$ = 12,000. Hou et al. (2022) proposed a novel Unet-LSTM network, achieving temporally sequenced hydrodynamic predictions of two-dimensional pressure and velocity characteristics for an underwater vehicle. However, the LSTM network necessitates transforming two-/three-dimensional image data into a one-dimensional sequence, leading to insufficient and potentially distorted spatial information representation. The Convolutional LSTM (ConvLSTM) network (Shi et al., 2015) addresses this by employing two-/three-dimensional image data as network input directly, integrating convolutional operation within a modified LSTM framework to extract physical field features. Compared with LSTM, ConvLSTM is more suitable for spatial sequence data processing and demonstrates superior performance in capturing spatiotemporal relationships. For instance, Han et al. (2019) combined the CNN and ConvLSTM to predict instantaneous pressure and velocity fields by giving sequential flow states at historical time steps for stationary cylinder and airfoil at unsteady wake flows. Moreover, ConvLSTM has been also successfully applied to turbulence prediction (Mohan et al., 2019) and fluid dynamics prediction (Wang et al., 2024). The aforementioned models are aimed at the dynamic flow field, however, lack the ability to identify the dynamic boundary of the moving structure that must exists in the FSI system.

The second type is focused on the unsteady fluid with moving boundary, which is limited to accurately predicting the transient structural dynamics or displacement at future occasions in a quantitative form. For cylindrical vibration, Cheng et al. (2021) combined physics-informed neural network (PINN) and LSTM to predict fluid force, oscillation velocity, and displacement for VIV of the cylinder. For aeroelastic airfoil, with Mach number, pitching angle, and plunging displacement as input variables, Li et al. (2019) proposed an LSTM-based unsteady aerodynamic reduced order model (ROM) for NACA 64A010 flutter airfoil to forecast corresponding time-dependent lift and moment coefficients, along with pitching and plunging responses. Similar research for flutter airfoils can be found in Refs. (Kou & Zhang, 2021; Li et al., 2021; Liu et al., 2023). For high-lift airfoil realized by mechanically variable camber, Zhang et al. (2021) predicted unsteady aerodynamic coefficients based on the force factors at the historical moment and deformation information at the current moment by utilizing the LSTM network. Above unsteady fluid models are capable of directly yielding the dynamic structural response of an oscillating or deformable rigid with moving boundaries, thereby encompassing characteristics of FSI systems to some extent. Nevertheless, these models fall short in capturing and leveraging the instantaneous state of the flow field, an aspect crucial for comprehensive FSI analyses.

To sum up, most prevailing time series deep models only consider either structural responses or flow field states in addressing unsteady fluid dynamics and FSI challenges by treating them as inputs and outputs within their network architectures and rarely involve the combination of these two categories. However, the inherent complexity of the actual FSI process necessitates consideration of flow field features, structure responses, and their interaction in each time step. In accordance with the principle of FSI, several spatial-temporal deep models for FSI based on this bidirectional coupling route have been proposed recently. Nava et al. (2022) interacted the IBM with pressure and velocity fields, trained via a Unet architecture, to establish an accurate FSI modeling system for a soft robot, focusing on simulating a single time step. Han et al. (2022) combined DNN-based fluid model and structural dynamic equation by transmitting cylindrical surface pressure and structural responses at each time step, aiming to establish a precise and efficient dynamic boundary model for VIV of an isolated cylinder with single-degree-of-freedom (1DOF) at $Re$ = 100. Among them, the fluid model is composed of CNN and ConvLSTM. The inadequacy of this model is that the complicated multi-cylinder projects and multi-freedom problems were not involved. For VIV of a single cylinder with two degrees of freedom (2DOF), Rachit & Rajeev (2022) adopted the CRAN framework and POD-RNN framework to build the flow field evolution model and structural motion model, respectively. The above two deep models constitute the resultful spatial-temporal model of VIV systems. Not limited to rigid body motion, Fan & Wang (2024) proposed a novel differentiable hybrid neural network that embeds partial differential equations into a deep-model architecture for FSI modeling of a rigid cylinder and a flexible plate. More specifically, this is achieved through the training of a ConvLSTM flow network integrated with computations of solid dynamics, realizing sequence-to-sequence prediction. Above FSI modeling generally applied a single bluff or elastic body, have not involved to the multiple moving objects with complex interaction between each other.

As a typical representative of FSI involving multiple moving objects, multiple cylinders undergoing VIV have been extensively studied (Borazjani & Sotiropoulos, 2009; Lin et al., 2013; Griffith et al., 2017). In this paper, we construct a FSI deep model for the VIV of two elastically mounted cylinders in tandem arrangement at $Re$ = 200. Double oscillators exhibit increased subtlety in temporal dynamics compared to isolated ones, manifested in more complicated flow patterns, larger vibrating amplitudes, and lift force (Borazjani & Sotiropoulos, 2009). This is closely related to the interaction between the double cylinders in relative proximity. The motion state of the downstream cylinder will be affected by the flapping of the shedding vortex from the upstream cylinder, which then feeds back into the flow field around the upstream one, resulting in either positive or negative coupling. Consequently, the nonlinear multi-boundary movement and complicated interacting flow fields amplify the complexity associated with modeling FSI systems. Current FSI network architectures utilize CNNs to predict physical spatial behaviors. However, generally CNNs with image-to-image frameworks absent global regularization and are less liable to precisely locate objects within image data. These drawbacks hinder the accurate reconstruction or prediction of complex global physical systems, especially for FSI projects with multiple dynamic boundaries. Moreover, existing networks that integrate CNNs with LSTMs or ConvLSTMs lack the connection between feature dimensionality reduction and expansion. Unet architecture bridges the encoder and decoder (Ronneberger et al., 2015), enabling integration of input image information from dimensionally reduction stages into the ascending dimension features, contributing to restoring the feature loss caused by the encoder in dimensionality reduction sampling. In contrast with CNNs, the Fourier neural operator (FNO), first presented by Li et al. in 2021, is able to learn parametric mappings in Fourier Spaces rather than Euclidean Spaces, leveraging the favorable inherent properties of function Spaces for enhanced performance (Li et al., 2021). Additionally, FNO demonstrates excellent proficiency in learning parametric partial differential equations, meaning its relevance with spectral methods and heightened suitability for physical field reconstruction and prediction tasks (Xiao et al., 2024; Wen et al., 2022; Li et al., 2023; Gopakumar et al., 2023). Crucially, FNO excels in capturing high-frequency information, with the oscillating cylinder edge representing a high-frequency region. This property endows FNO with superior ability in forecasting dynamic boundaries.

Specifically, with regard to VIV of multi-cylinder, considering interactionally coupled dynamic boundary prediction of spatial-temporal evolution, we construct an innovative network that includes the FNO module and U-shaped framework with ascending and descending dimension characteristics and named it USFNO in this paper. The USFNO has stronger physical field forecasting performance rather than CNNs. To leverage the physical features of the historical flow field effectively, a reasonable length of historical time series is utilized as the modeling input, with information extraction by a ConvLSTM fused with temporal information (FConvLSTM). Simultaneously, a novel spatial-temporal deep model named USFNO-FConvLSTM is proposed in this work combining the physical prediction ability of USFNO with the spatiotemporal prediction capability of FConvLSTM. Further, an innovative USFNO-based hybrid neural model is constructed for predicting FSI dynamics. This FSI neural model consists of USFNO-based FSI recurrent units that contains three essential elements: flow field evolution, structural response, and coupling interaction between them. Therein, the time evolution of the flow fluid with cylindrical movement information is realized by USFNO-FConvLSTM, effectively tackling the time-consuming issue of iterative flow-field computations encountered in FSI processes. Structural responses for multiple cylinders are computed iteratively by the structural motion control equations. The most critical coupling interaction is accomplished by the surface force transmission on cylinders. In the proposed FSI model, USFNO-FConvLSTM is applied to predict the flow field state at the next time step based on the historical multi-time step and current structural response. Focusing on multiple time steps enhances the forecast accuracy for future flow fields, which constitutes the advantage of the model distinguishing from the existing FSI neural models. Yin et al. (2022) recorded that the wall shear stress can be omitted under the high Reynolds number flows. However, the laminar flow at low $Re = 200$ is considered in this paper. Therefore, different from the previous FSI neural networks that solely consider surface pressure (Han et al., 2022; Rachit & Rajeev, 2022), this paper introduces wall shear stress as another non-negligible component of the transmitted force. Besides that, a time series model of wall shear stress constructed by MLP is juxtaposed with flow field modal and structural solver to constitute the novel USFNO-based hybrid neural model. The proposed FSI model allows sequence-to-point training and prediction.

Focusing on the VIV of two cylinders in tandem arrangement with 1DOF, firstly, we apply the dynamic mesh technology to generate high-precision simulation data that are used as network modeling. Then, the FSI network model with high predictive performance is established through parameter and contrast tests, as follows. Our work involves the effects of sequence length on the modeling accuracy of physical fields. Furthermore, it is demonstrated that the USFNO-FConvLSTM is superior to the CNN-FConvLSTM and Unet-FConvLSTM in predicting flow field state under various modeling parameters. In addition, the excellent performance of the proposed FSI neural network is confirmed through appropriate experiments on VIV of both isolated and double cylinders. Concretely, we compare its corresponding prediction accuracy and reliability reflected in flow field evolution and structural responses with other FSI models. In conclusion, this paper makes the following novel contributions:

(1) A novel FSI network framework consisting of the fluid dynamic model, structural dynamic solver, and coupling interaction, is developed to accommodate the nonlinear flow field evolution and structural motion responses, yielding an accurate and reliable FSI model, i.e., USFNO-based hybrid neural network model.
(2) We propose a novel spatial-temporal model allowing sequence-to-point flow field prediction based on the composition of FNO, FConvLSTM, and U-shaped framework, named USFNO-FConvLSTM, which can greatly capture the multiple dynamic boundaries of vibrating bluff bodies.
(3) An accurate time series model of wall shear stress for the dynamic boundary is introduced as part of the FSI model to perfect the force transfer process of the coupling interface.
(4) Two cylinders in tandem arrangement with nonlinear multi-boundary movement and complicated interacting flow fields are applied to analyze and verify the excellent modeling property of the proposed FSI neural network framework.

The structure of the paper is outlined as follows: The framework of the proposed FSI deep model is described in Section 2. Problem statement and data acquisition are progressively elaborated upon in Section 3. Section 4 presents the modeling accuracy and reliability of the established FSI model. Conclusions are summarized in the end.

## 2. The framework of fluid-structure interaction surrogate model

The traditional FSI numerical simulation framework comprises a fluid dynamic solver and a structural dynamic solver. In this paper, we propose a novel FSI neural model and name it USFNO-based hybrid neural network, as shown in Figure 1, which refers to the traditional FSI numerical solver and is trained by the spatial-temporal time sequence data calculated from its numerical simulation. Taking into consideration the time-consuming CFD computations and the programmability of structural dynamics for high-dimensional FSI solver, we establish a FSI hybrid neural framework consisting of two parallel-coupled solvers: fluid deep learning solver and structural dynamic solver, as described in Figure 1(a). In contrast to conventional FSI simulators that rely on a single time-step iterative approach, the proposed FSI hybrid neural network forecasts the current flow field state ($F_i$) by leveraging the previous $N$ flow field states ($F_{i-N},…,F_{i-1}$) along with the structural response from the prior moment ($S_{i-1}$). Additionally, the above two solvers execute information exchange and nonlinear coupling through the surface fluid force (named as $F_b$ calculated by surface pressure '$P$' and wall shear stress '$WS$') and structural displacement (denoted as '$y$') in the process of time-advancing iteration. The overall hybrid neural model, which is formulated as a ConvLSTM sequence network, consists of USFNO-based FSI recurrent units connected by temporal features ('Hidden State' in Figure 1(a)). The detailed schematic and essential features of this novel proposed USFNO-based FSI recurrent unit is illustrated in Figure 1(b). Concretely, recurrent unit includes the fluid deep learning model, structural dynamic equations, and their data transfer. Specifically, to be consistent with the principle of traditional FSI solver so as to ensure the solving accuracy of fluid dynamics, the fluid deep learning model is composed of two separate data-driven solvers for wall shear stress and flow field subdomains, which correspond to two independent neural networks (i.e, MLP and USFNO-FConvLSTM).

The innovation and application of the USFNO-based hybrid neural network, along with its constituent FSI recurrent unit, can significantly diminish the computational time associated with FSI, while concurrently enhancing predictive accuracy in comparison to existing FSI surrogate models. The detailed architecture of the fluid, solid, and their coupling components of the USFNO-based FSI recurrent unit is elaborated in Section 2.1 to Section 2.3, respectively, as described below.

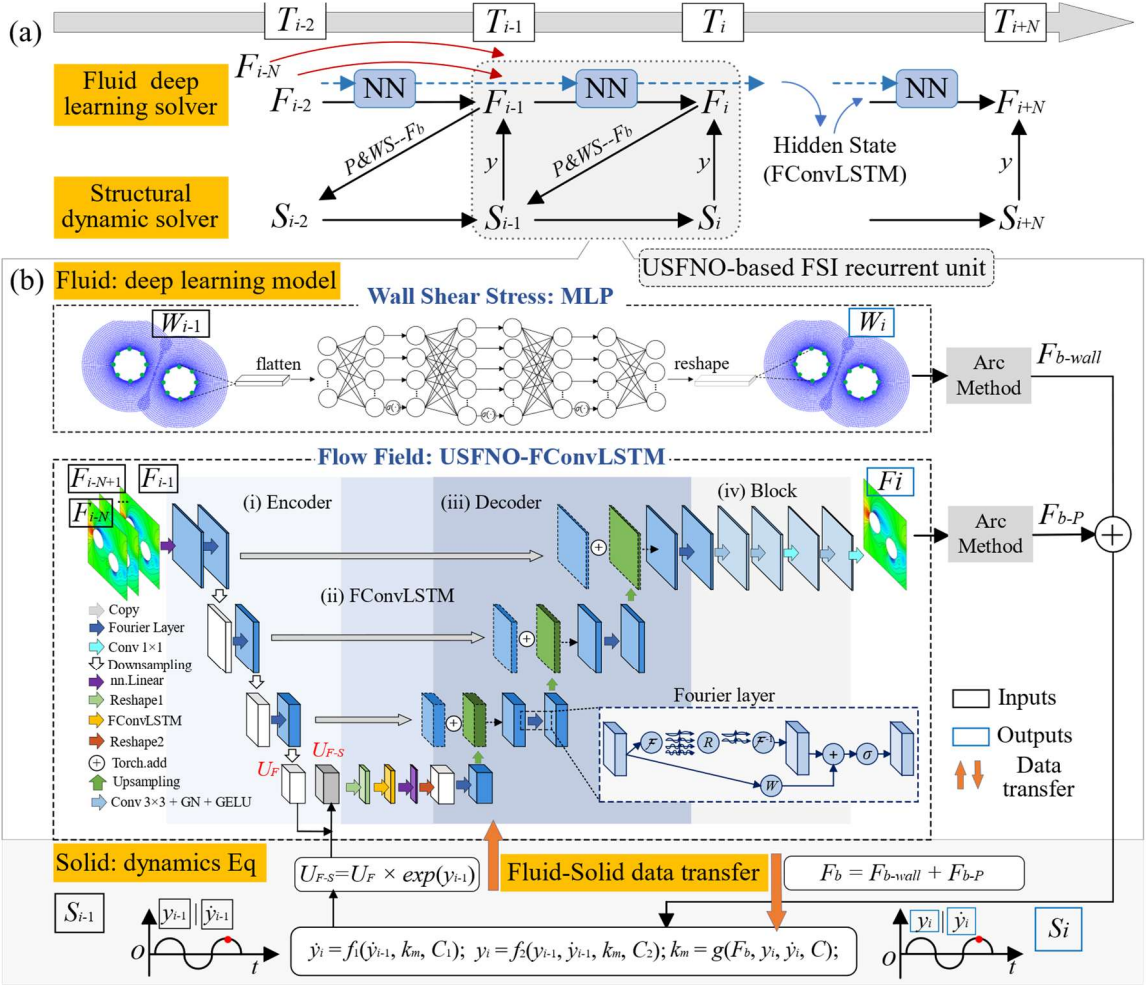

**Figure 1.** (a) Overview framework of USFNO-based hybrid neural network for FSI simulation (time-advancing iteration between fluid deep learning solver and structural dynamic solver), where '$F_i$' and '$S_i$' represent the flow fluid and structural dynamics at transient time step '$T_i$', respectively; 'NN' denotes the trainable fluid deep learning model; The overall hybrid neural model is characterized as a ConvLSTM sequence network which is consists of USFNO-based FSI recurrent units (b) The detailed architecture of the USFNO-based FSI recurrent units, which includes the fluid deep learning model, structural dynamic equations, and their data transfer; Model inputs and outputs are marked as black and blue, respectively.

### 2.1. Fluid deep learning model

Within the FSI analyses, in consideration of the substantial computational time expended by CFD solvers relative to the expedited derivation of structural responses, we construct a fluid deep learning model as a substitution to accelerate the solving speed for the fluid part on the basis of retaining the structural calculation principle. As shown in Figure 1(b), this novel fluid deep learning model consists of two separate data-driven solvers for wall shear stress and flow field submodules.

#### 2.1.1. Wall shear stress: MLP

The friction force caused by viscous fluid on the cylindrical surface, which appears in the form of wall shear stress, is another pivotal component of the periodic lift and drag of an oscillator in addition to the wall pressure. The detailed steps for integrating the above two discrete components into the forces are described in Section 3.3. Yin et al. (2022) recorded that the wall shear stress can be omitted under the high Reynolds number flows. However, the laminar flow with $Re$ = 200 considered in this paper will induce Reynolds shear stress that has a non-negligible contribution for periodic forces, which is demonstrated in Section 3.3. As depicted in Figure 1(b), the wall shear stress module adopts MLP to realize the progressive prediction of time step from input $t_{i-1}$ to output $t_i$ corresponding to the discrete stress from $W_{i-1}$ to $W_i$ with a single dimension, respectively.

$$W_i = f^w(W_{i-1}). \qquad (1)$$

$$W_i = \{w_i^1, w_i^2, \ldots, w_i^n, \ldots, w_i^N\}. \qquad (2)$$

where $f^w$ is the trainable function relationship; $w_i^n$ denotes the $n$th predictive stress point at $t_i$; and superscript $N$ and subscript $i$ are the discrete point number of shear stress and time step $t_i$, respectively. Furthermore, the hidden layers are composed of four fully connected layers with 512, 1024, 1024, and 512 neurons in each layer. To improve the convergence speed and solve the vanishing gradient phenomenon of this deep neural network, the Gaussian Error Linear Unit (GELU) is inserted at intervals. The L1 loss function $\mathcal{L}_{\text{stress}}$ is utilized to optimize the network parameter of this MLP framework:

$$\mathcal{L}_{\text{stress}} = \frac{1}{N} \sum_{n=1}^{N} |\hat{w}_i^n - w_i^n|. \qquad (3)$$

where $\hat{w}_i^n$ denotes the $n$th labeled (truth) stress point at $t_i$. Subsequently, integrating the wall shear stress in $W_i$ can obtain the wall shear force $F_{b\text{-wall}}$ at $t_i$. The integration arc method is described in Section 3.3. It is noteworthy that the obtained $F_{b\text{-wall}}$ constitutes one of the components of the interaction forces facilitating FSI.

*2.1.2. Flow field: USFNO-FConvLSTM*

Flow field evolution can be modeled using the novel proposed USFNO-FConvLSTM architecture, as described in Figure 1(b). The input of USFNO-FConvLSTM involves the historical sequential flow field with $N$ steps ($t_{i-N} \sim t_{i-1}$), subsequently, deriving the predicted flow field at $t_i$:

$$F_i = f^F(F_{i-1}, F_{i-2}, \ldots, F_{i-N}, S). \qquad (4)$$

where $f^F$ represents the learnable functional relationship; $S$ denotes the structural responses and will be explained in Section 2.2. Emphatically, USFNO-FConvLSTM is classified as the prediction model of multiple time sequence and is an improvement on the traditional and common single-time step iteration. It should be noted that $[F_{i-N}, \ldots, F_i]$ is the discrete two-dimensional matrix flow field data, and the cylinder is filled with zeros, which will be explained in detail in the Section 3.3. To extract more valuable spatiotemporal features from the flow field, the USFNO-FConvLSTM network is structured in a U-shaped architecture comprising four submodules: Encoder, FConvLSTM, Decoder, and Block (in Figure 1(b)). Among them, the encoder and decoder are composed of convolution Fourier layers (Conv-Fourier layer) and deconvolution Fourier layers (DeConv-Fourier layer), respectively. Consequently, the USFNO-FConvLSTM architecture depicted in Figure 1(b) can be represented in an alternative form, as illustrated in Figure 2, it entails the following:

Transfer the input $v(x) = [F_{i-N}, \ldots, F_{i-1}]$ to a representation space $z_{l_0}(x)$ with the more channels $d_c$, as follows:

$$z_{l_0}(x) = P_1(v(x)). \qquad (5)$$

where $P_1$ represents a linear layer. The output results $z_{l_0}(x)$, as the initial input to the Fourier layer, can be regarded as $N$ 2-D images with $d_c$ channels. Therein, applying $L$ Conv-Fourier layers for the Encoder and $M$ DeConv-Fourier layers for the Decoder. The output of Encoder and Decoder is $z_{l_L}^{\mathcal{C}}$ and $z_{m_M}^{\mathcal{D}}$, respectively. The superscript $\mathcal{C}$ and $\mathcal{D}$ represent the decrease and increase in image resolution. The stacked Fourier layers and FConvLSTM $z_T$ constitute an iterative architecture $I$, corresponding to the Figure 2(a).

$$I: z_{l_1}^{C_1} \rightarrow z_{l_2}^{C_2} \rightarrow \ldots \rightarrow z_{l_L}^{C} \rightarrow z_T \rightarrow z_{m_1}^{D_1} \rightarrow z_{m_2}^{D_2} \rightarrow \ldots \rightarrow z_{m_M}^{D}. \tag{6}$$

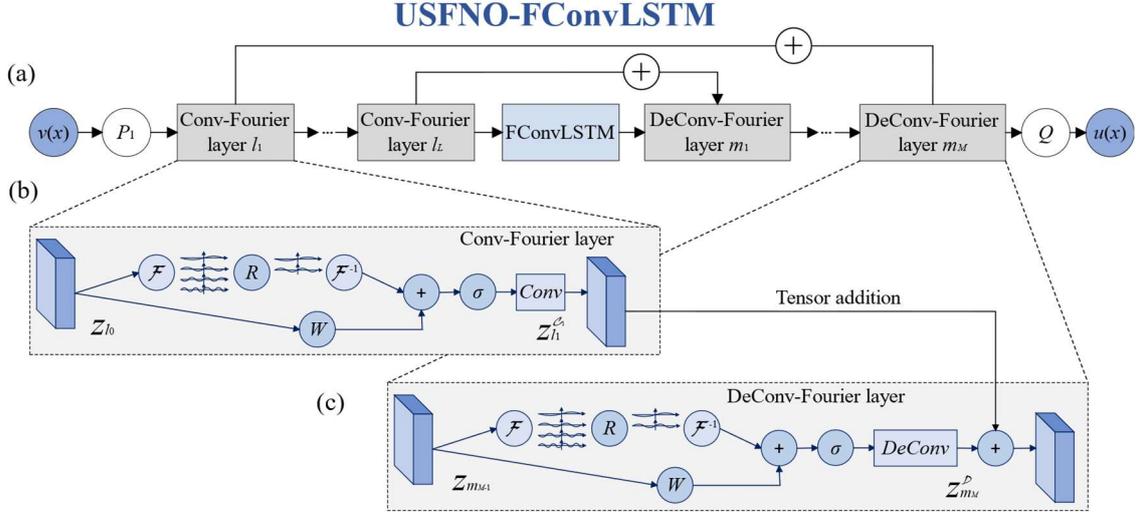

**Figure 2.** Architecture of USFNO-FConvLSTM. (a) the full USFNO-FConvLSTM architecture; where $v(x)$ represents the input flow field data; $P_1$ is a linear layer that increases the channels of input flow field; followed by the Conv-Fourier layers, FConvLSTM and DeConv-Fourier layers; $Q$ is a convolution and activation operation that reduces the channels of target flow field $u(x)$ (b) Conv-Fourier layer, where $\mathcal{F}$ and $\mathcal{F}^{-1}$ denotes the Fourier transform and inverse Fourier transform, respectively; where $R$ and $W$ are weight matrixes, respectively; and $\sigma$ is the activation function; '*Conv*' represents the convolution operation (c) DeConv-Fourier layer: '*DeConv*' denotes the deconvolution operation. 'Tensor addition' bridge the encoder and decoder modules.

**Conv-Fourier layer:** As shown in Figure 2(b), each Conv-Fourier layer contains a Fourier layer and a convolution layer (down sampling operation $\mathcal{C}$). The update $z_{l_j}^{C_j} \mapsto z_{l_{j+1}}^{C_{j+1}}$ can be conducted through $l_{j+1}$ layer.

Each Fourier layer performs a Fourier transform (FFT) (Li et al., 2021) and its inverse transform (IFFT) that are respectively represented by $\mathcal{F}$ and $\mathcal{F}^{-1}$, meanwhile, entails a weight matrix $R$:

$$\mathcal{K}(z_{l_j}^{C_j})(x) = \int k(x,y) z_{l_j}^{C_j}(y) dz_{l_j}^{C_j}(y) = \mathcal{F}^{-1}(R_{l_j} \cdot \mathcal{F}(z_{l_j}^{C_j}))(x). \tag{7}$$

where the non-local integral operator $\mathcal{K}$ and kernel $k(x,y)$ are both learnable and parameterized. And $\mathcal{K}$ is implemented by mapping the representation $z_{l_j}^{C_j}$ to the Fourier space. $R_l$ is the parameterization of the periodic function $k$ after the Fourier transform. Therein, the Fourier transform $\mathcal{F}$ is employed on each channel $d_c$ of $z_{l_j}^{C_j}$. To enhance trainable efficiency, only the first $k_{max}$ Fourier modes are utilized while truncating the remaining ones. In this paper, excluding the dimensions of samples $N$ and channels $d_c$, the input flow field $z_{l_j}^{C_j}$ can be regarded as a two-dimensional discretization matrix $[X_1, X_2]$ with the resolution $h \times w$, corresponding 2D inputs $k_{max} = k_1 \times k_2$. Where $k_1$ and $k_2$ denotes the number of Fourier models applied in dimension $X_1$ and $X_2$. Therefore, the FFT $\mathcal{F}$ and IFFT $\mathcal{F}^{-1}$ can be calculated as:

$$(\mathcal{F}(z_{l_j}^{C_j}))(k_1,k_2) = \sum_{x_1=0}^{h-1} \sum_{x_2=0}^{w-1} (z_{l_j}^{C_j})(x_1,x_2) e^{-2i\pi(x_1 k_1/h + x_2 k_2/w)} \tag{8}$$

$$(\mathcal{F}^{-1}(z_{l_j}^{C_j}))(x_1,x_2) = \sum_{k_1=0}^{h-1} \sum_{k_2=0}^{w-1} (z_{l_j}^{C_j})(k_1,k_2) e^{-2i\pi(x_1 k_1/h + x_2 k_2/w)} \tag{9}$$

After the operations of nonlinear activation function $\sigma$ and linear transformation weight $W$, the output of Fourier layer can be expressed as:

$$z_{l_{j+1}}^{\mathcal{C}_j}(x)=\sigma(\mathcal{K}(z_{l_j}^{\mathcal{C}_j})(x)+W_j \cdot z_{l_j}^{\mathcal{C}_j}(x)). \tag{10}$$

A convolution layer is then applied to reduce the image resolution and increasing the channel:

$$z_{l_{j+1}}^{\mathcal{C}_{j+1}}(x)=\mathcal{C}_{j+1}(z_{l_{j+1}}^{\mathcal{C}_j}(x)). \tag{11}$$

The output of $L$th layer is $z_{l_L}^{\mathcal{C}}$ viewed as an image ($n_x \times n_y$) with $d_{c_L}$ channels and $N$ batch-size (i.e., [$N$, $d_{c_L}$, $n_x$, $n_y$]).

The Fourier layer effectively captures the frequency components of the flow field by transforming the input data into the frequency domain, which is particularly advantageous for physical phenomena characterized by periodic and oscillatory behaviors, such as the cylindrical VIV studied in this study. Additionally, stacking multiple Fourier layers enhances the capacity for reconstructing physical fields. Moreover, the Conv-Fourier layer facilitates dimensionality reduction and information extraction from flow field data while preserving the benefits of Fourier layer, thereby significantly enhancing computational efficiency and generalization capabilities.

**FConvLSTM:** To better capture useful spatial features in temporal evolution, ConvLSTM was proposed by Shi et al. (2015) to enhance the performance of traditional LSTM (Hochreiter & Williamson, 1997). In principle, LSTM introduces cell states and memory mechanisms, so that provides superior performance of gradient continuity and long-term memory compared with traditional RNN. Additionally, it has better learning capacity and data processing ability for dynamic characteristics. Both two methods are composed of input gate $i_t$, hidden gate $f_t$, and output gate $o_t$. Among them, the LSTM cell specializes in processing time series data by transmitting the internal state through the fully connected layer. Hence, it requires limiting the input and hidden gates to a one-dimensional vector. That is, a two- or three-dimensional image or data field must be transformed into one-dimensional data, which destroys the authenticity of multidimensional data and complicates the capture of spatial characteristics. Nevertheless, the ConvLSTM is capable of directly utilizing two-/three-dimensional flow field data as input, integrating convolution operations within a modified LSTM framework to extract physis field features. This paper considers the two-dimensional flow field with cylindrical dynamic boundary, characterized by abundant spatial information and strong interaction correlations among the data points within the flow field. Consequently, in comparison to LSTM, ConvLSTM demonstrates a superior capability to handle spatio-temporal evolution characteristics and local dependencies within sequences while effectively preserving the positional information of data points in the flow field, as shown in Figure 3.

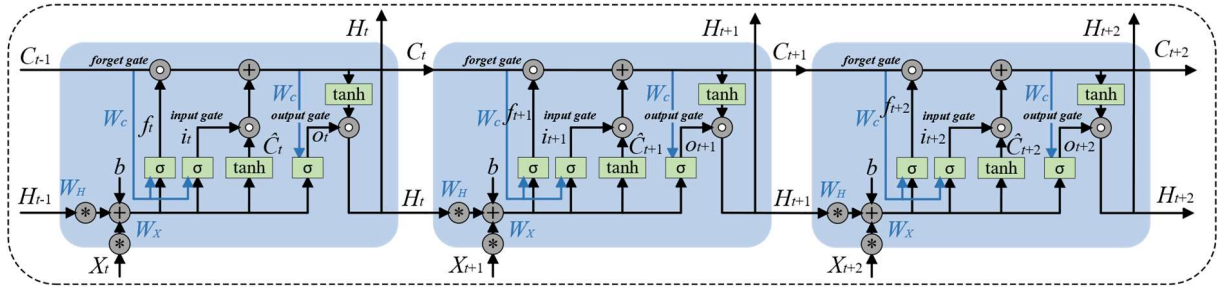

**Figure 3.** ConvLSTM architecture.

In time series prediction, for ConvLSTM, the sequence length of input can be longer than that of the output, which can realize the prediction from multi-time step (i.e., $t_{i-N} \sim t_i$) to future single time step $t_i$. The operational principle of ConvLSTM can be expressed as:

$$i_t = \sigma(W_{xi} * X_t + W_{hi} * H_{t-1} + W_{ci} \odot C_{t-1} + b_i), \tag{12}$$

$$f_t = \sigma(W_{xf} * X_t + W_{hf} * H_{t-1} + W_{cf} \odot C_{t-1} + b_f), \tag{13}$$

$$C_t = f_t \odot C_{t-1} + i_t \odot \tanh(W_{xc} * X_t + W_{hc} * H_{t-1} + b_c), \tag{14}$$

$$o_t = \sigma(W_{xo} * X_t + W_{ho} * H_{t-1} + W_{co} \odot C_{t-1} + b_o), \tag{15}$$

$$H_t = o_t \odot \tanh(C_t). \tag{16}$$

where $X$, $H$, and $C$ denote the cell inputs, hidden states, and cell outputs, respectively. Sigmoid function $\sigma$ is chosen as the activation function here; $W$ and $b$ represent the weights and biases for gates of $i_t$, $f_t$, $o_t$, respectively. Convolution operator '*' and Hadamard product '$\odot$' are used to accomplish the inner connection. Different from traditional ConvLSTM which realizes the convolution operation in channels, convolution calculation is carried out on the features of multiple time steps to realize the fusion of multiple time steps in this paper. We name this modified network time-fused ConvLSTM, i.e., FConvLSTM. This method is proved to have advantages over traditional ConvLSTM methods, as shown in Section 4.2.2. Specific steps entail as follows: switching the position of time step $N$ and channel $d_{c_L}$ to reshape the $z_{l_L}^{\mathcal{C}}$ as $[d_{c_L}, N, n_x, n_y]$ by introducing a reshape function $R_1$. Then, employing the 'FConvLSTM' function $\mathcal{T}$ to conduct time-sequence training and capture physical features:

$$z_{l_L}^{\mathcal{C}\prime}(x) = \mathcal{T}(R_1(z_{l_L}^{\mathcal{C}}(x))). \tag{17}$$

The linear layer $P_2$ fuses the comprehensive features of multiple historical time steps into a single future time step (i.e., $[d_{c_L}, 1, n_x, n_y]$). Furthermore, second reshape function $R_2$ is utilized to transform the output shape of $[1, d_{c_L}, n_x, n_y]$, obtaining the output of the time series part $Z_T$.

$$z_T(x) = R_2(P_2(z_{l_L}^{\mathcal{C}\prime}(x))). \tag{18}$$

**DeConv-Fourier layer:** Similar to the architecture set in Conv-Fourier layer, each DeConv-Fourier layer has a Fourier layer and a deconvolution layer (upsampling operation $\mathcal{D}$), as shown in Figure 2(c). The inverse convolution operation is utilized to increase the image resolution and decrease the channel. The output of the $(k+1)$th layer $z_{m_{k+1}}^{\mathcal{D}_{k+1}}$ can be expressed as:

$$z_{m_{k+1}}^{\mathcal{D}_{k+1}}(x) = \mathcal{D}_{k+1}(\sigma(\mathcal{F}^{-1}(R_{m_k} \cdot \mathcal{F}(z_{m_k}^{\mathcal{D}_k}))(x) + W_k \cdot z_{m_k}^{\mathcal{D}_k}(x))). \tag{19}$$

In Figure 1(b) and Figure 2, we conduct the U-shaped framework to bridge the encoder and decoder modules. Specifically, we integrate the last time step feature of $z_{l_{j+1}}^{\mathcal{C}_{j+1}}$ and corresponding DeConv-Fourier layer $z_{m_{k+1}}^{\mathcal{D}_{k+1}}$ by performing the tensor-addition principle, as shown by the gray arrow in Figure 1(b). The output of the decoder module is $z_{m_M}^{\mathcal{D}}$.

To improve fitting accuracy, two modules both consisting of appropriate convolutions and activations are introduced to decrease the target channel number.

$$F_i = u(x) = Q(z_{m_M}^{\mathcal{D}}(x)). \tag{20}$$

$$\mathcal{L}_{\text{Field}} = \frac{1}{N} \sum_{n=1}^{N} |\hat{F}_i^n - F_i^n|. \tag{21}$$

where $\widehat{F}_i$ is the labeled flow field at $t_i$; $N$ corresponds the number of discrete points in the flow field. The L1 loss function $\mathcal{L}_{\text{Field}}$ is utilized to optimize the network parameter.

The fluid dynamic $F_{b\text{-}P}$ is calculated from the predicted pressure field $F_i$. Then, $F_{b\text{-}wall}$ and $F_{b\text{-}P}$ are added to obtain the total flow field force $F_b$ of the fluid part participating in the fluid-structure coupling. $F_b$ is the exciting force of the structural motion control.

### 2.2. Solid dynamic equations

The cylindrical VIV can be regarded as a stiffness-damping system. Its structural dynamics equation is written as:

$$m_{cyl}\ddot{y}+c\dot{y}+ky=F_b. \tag{22}$$

where $m_{cyl}$ denotes the oscillating mass of cylinder; $c$ and $k$ represent the damping and spring stiffness acting on the oscillator, respectively. $F_b$ can facilitate the position migration of the cylinder from $t_{i-1}$ to $t_i$. As expressed in Figure 1(b), the structural dynamics equation can be discretized and solved by fourth-order Runge-Kutta method:

$$k_m = g(F_b, y_{i-1}, \dot{y}_{i-1}, C), \tag{23}$$
$$\dot{y}_i = f_1(\dot{y}_{i-1}, k_m, C_1), \tag{24}$$
$$y_i = f_2(y_{i-1}, \dot{y}_{i-1}, k_m, C_2). \tag{25}$$

where $k_m$ represents the variable coefficients of fourth-order Runge-Kutta. $g$, $f_1$ and $f_2$ are the corresponding functions. $\dot{y}$ and $y$ denote the instantaneous vibration velocity and displacement, respectively, which can be derived from the corresponding structural responses at $t_{i-1}$. Corresponding $C$, $C_1$ and $C_2$ are the other key parameters. It is worth noting that the time evolution curves of the moving structure responses, characterized by quantitative values, can be derived through an iterative solution of the discrete equations (Equation (22) - Equation (25)), reflecting how the FSI neural model processes the moving structure.

### 2.3. Fluid-Solid data transfer

The physical data transfer between the flow field and the solid structure is embodied by two orange arrows in Figure 1(b).

First is the feedback of the structure to the flow field. The feedback $U'_F$ of structural oscillation to the flow field evolution is realized by fusing the structural responses ($y_{i-1}$) at $t_{i-1}$ with the low-dimensional fluid features $U_F$ obtained by the high-dimensional reduction at $t_{i-1}$, as following:

$$U'_F = U_F * \exp(y_{i-1}). \tag{26}$$

Ulteriorly, the flow field state ($F_i$) at the next moment $t_i$ can be obtained by combining the historical flow field information (i.e., $F_{i-N} \sim F_{i-1}$) and structural response $y_{i-1}$ (corresponding to the boundary conditions in CFD solver):

$$F_i = f^F(F_{i-1}, F_{i-2}, \ldots, F_{i-N}, y_{i-1}). \tag{27}$$

Because the moving structure interior in the flow field is filled with zero values, the structure position in the flow field image can be directly predicted through the fluid deep learning model, which reflects how the FSI neural model handles moving structures. The presence of zero-value regions within the moving structure can result in pronounced gradient features at the boundaries of the structure and the surrounding flow field. The FNO employed in this study is particularly adept at capturing high-frequency information, thereby exhibiting enhanced predictive capabilities for dynamic boundaries characterized by such traits.

Second is the force information transfer of the flow field to the structure. Combined with Figure 1(b) and the previous analysis, surface fluid force $F_b$ can be obtained by integrating discrete surface wall shear stress $W_i$ and pressure $F_i$ along the cylindrical periphery:

$$F_b = F_{b\text{-}P} + F_{b\text{-}wall}. \tag{28}$$

Introducing the $F_b$ into the discretized structural dynamics equations (Equation (22) - Equation (25)) can derive the structural response information ($y_i$) at the future state $t_i$, as explained in Section 2.2.

Continuous iteration can complete the training and prediction of the whole FSI network.

## 3. Problem statement and data acquisition

VIV is a typical phenomenon of FSI (Khalak & Williamson, 1997). In some sea conditions, significant VIV will occur in marine risers, especially for multiple riser systems. The complex interaction between the upstream and downstream cylinders will aggravate the risers. And with the advance of time, the risers will produce fatigue damage. Therefore, it is essential to establish an appropriate deep model considering of time effect for multiple cylinders. In this paper, two tandem cylinders with the center-to-center spacing ratio of 1.5 under 1DOF are specified as the application object to verify the validity and superiority of the proposed USFNO-based hybrid neural network. Two cylinders have the same inherent attribute. To prove the generalization of modeling, an isolated cylinder with 1DOF is introduced as the other physical example.

*3.1. Vortex-Induced Vibration*

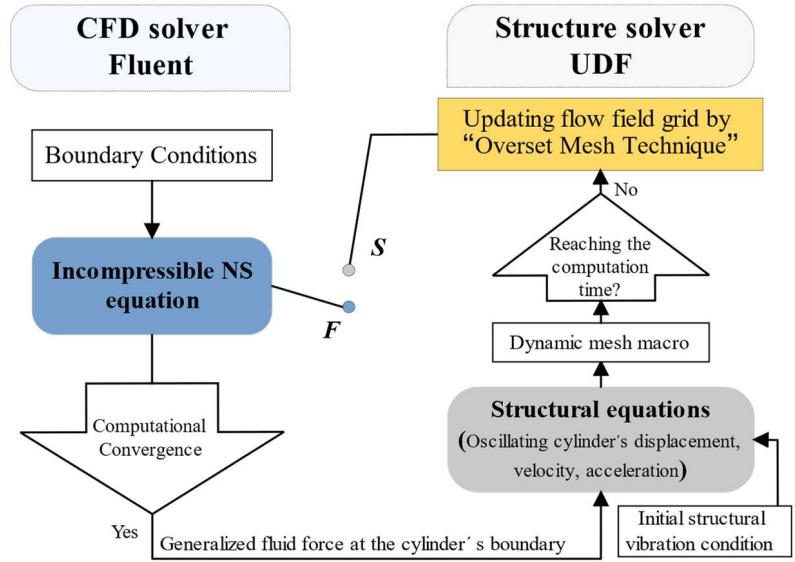

**Figure 4.** Schematic diagram of numerical simulation for bidirectional FSI of the elastically mounted cylinder.

In this paper, the commercial CFD software (i.e., Fluent, with the pressure-based solver) and structural dynamic principle, combined with the user-defined function (UDF) and the overset mesh technique (Xie et al., 2019) are used to establish the VIV numerical model of two elastically mounted cylinders. The specific schematic diagram of this bidirectional FSI is shown in Figure 4.

For fluid mechanics, two-dimensional unsteady incompressible laminar flow can be governed by the incompressible Navier-Stokes equation as follows:

$$\begin{cases} \partial u_i/\partial x_i = 0 \\ \partial u_i/\partial t + \partial(u_i u_j)/\partial x_j = -\partial P/\partial x_i + \partial^2 u_i/Re\partial x_i \partial x_j \end{cases} \quad (29)$$

where $x_i$ and $u_i$ are, respectively, the Cartesian coordinates and corresponding instantaneous velocity components. Here, $P$ and $Re$ are pressure and Reynolds number ($Re$), respectively. The latter is defined as:

$$Re = U_0 D \rho / \mu. \quad (30)$$

where $U_0$ is the incoming flow velocity; $D$ is the diameter of the oscillator; $\rho$ denotes the fluid density; and $\mu$ represents the dynamic coefficient of viscosity. The pressure and velocity distributions within the outflow field, as well as on the two-dimensional cylindrical surface, can be derived from the specified boundary conditions and Navier-Stokes equations. Additionally, the fluid lift force ($F_L$) exerted on the cylindrical boundary can be extracted from flow field data, representing the direct force applied by the fluid to induce structural response, as illustrated in Figure 4.

For structural motion, the elastic oscillator can be equivalent to the mass-damping-spring system, and its motion equation can be expressed as:

$$\begin{cases} m_{cyl}\ddot{y}_1 + c\dot{y}_1 + ky_1 = F_{L1} \\ m_{cyl}\ddot{y}_2 + c\dot{y}_2 + ky_2 = F_{L2} \end{cases}. \quad (31)$$

where $y_1$ and $y_2$ are respectively the vibration displacement of upstream and downstream cylinders perpendicular to the flow direction. $\dot{y}$ and $\ddot{y}$ denote the vibration velocity and acceleration, respectively. The lift forces in the $y$-direction for two cylinders are expressed as $F_{L1}$ and $F_{L2}$, respectively, corresponding to the $F_b$ shown in Figure 1. These two forces can be calculated by integrating the pressure field and wall shear stress exporting from the CFD, as shown in Figure 4. The specific integration process can be found in Section 3.3.

$$\begin{cases} F_{L1} = 0.5 C_{L1} \rho U_0^2 D \\ F_{L2} = 0.5 C_{L2} \rho U_0^2 D \end{cases}. \quad (32)$$

where $C_L$ is the corresponding lift coefficient. Another expression of governing equation can be obtained by substituting Equation (32) into Equation (31):

$$\begin{cases} \ddot{y}_1 + 4\pi \zeta f_{ny} \dot{y}_1 + 4\pi^2 f_{ny}^2 y_1 = C_{L1} \rho U_0^2 D / 2 m_{cyl} \\ \ddot{y}_2 + 4\pi \zeta f_{ny} \dot{y}_2 + 4\pi^2 f_{ny}^2 y_2 = C_{L2} \rho U_0^2 D / 2 m_{cyl} \end{cases}. \quad (33)$$

Thereinto, natural frequency of oscillator $f_{ny}$ is defined as $f_{ny} = \sqrt{k/m_{cyl}}/2\pi$; damping ratio is written as $\zeta = c/(2\sqrt{k/m_{cyl}})$. Utilizing the fluid forces and the structural vibration conditions from the preceding time step, structural motion equations (Equation (33)) are solved to derive the displacement, velocity, and acceleration of the cylinder's centroid. It is noteworthy that this oscillatory motion is implemented and updated by embedding the structural motion equation within the UDF of CFD, where the motion equations are discretized using the fourth-order Runge-Kutta method. The detailed discrete equation in $y$-direction is expressed as:

$$\begin{cases} \dot{y}_{t+1} = \dot{y}_t + (k_1 + 2k_2 + 2k_3 + k_4) dt/6 \\ y_{t+1} = y_t + dt \cdot \dot{y}_t + (k_1 + k_2 + k_3) dt^2/6 \end{cases} \quad (34)$$

$$\begin{cases} k_1=F_L/m_{cyl}-4\pi\zeta f_{ny}\dot{y}_t-4\pi^2 f_{ny}^2 y_t \\ k_2=F_L/m_{cyl}-4\pi\zeta f_{ny}(\dot{y}_t+0.5k_1 dt)-4\pi^2 f_{ny}^2(y_t+0.5\dot{y}_t dt) \\ k_3=F_L/m_{cyl}-4\pi\zeta f_{ny}(\dot{y}_t+0.5k_2 dt)-4\pi^2 f_{ny}^2(y_t+0.5\dot{y}_t dt+0.25k_1 dt^2) \\ k_4=F_L/m_{cyl}-4\pi\zeta f_{ny}(\dot{y}_t+k_3 dt)-4\pi^2 f_{ny}^2(y_t+\dot{y}_t dt+0.5k_2 dt^2) \end{cases} \quad (35)$$

where $k_1$, $k_2$, $k_3$, and $k_4$ represent the variable coefficients of fourth-order Runge-Kutta, corresponding to the $k_m$ shown in Figure 1; $dt$ is the time step of the FSI process; and subscript of $t$ in $y_t$ denotes the specific moment. The displacement and velocity of the centroid at the cylinder boundary are conveyed to the CFD solver through the dynamic mesh "DEFINE" macro within the UDF procedure. Subsequently, the flow field mesh, representing the movement of cylinders, is updated by integrating the overset mesh technique with the transient displacement and velocity of the cylinders. A detail implementation of overset mesh technique is provided in Section 3.2).

*3.2. Model description and VIV modeling*

In the current study, two VIV simulations, i.e., two cylinders in tandem arrangement and an isolated cylinder, are configured at $Re = 200$, each with 1DOF perpendicular to the flow direction. All of the oscillators set with the equal diameter of $D = 0.01$m are arranged in an unsteady laminar flow with reduced velocity $U_r = U_0/f_{ny}D$, density $\rho$ and viscosity coefficient $\mu$ of 5, 998.2 kg/m$^3$ and 0.001003 kg/(m·s), respectively. The mass ratio $m^*$ and natural frequency $f_{ny}$ of the isolated cylinder are 10 and 2.51327, respectively. These two attributes have the values of 2.5465 and 0.40192, respectively, for tandem cylinders. To stimulate high oscillating amplitude, the structural damping ratios of all oscillators are fixed as zero, i.e., $\zeta = 0$.

$$m^* = m_{cyl}/0.25\pi\rho D^2. \quad (36)$$

As drawn in Figure 5(a,b), the rectangular computational domains of an isolated oscillator and two oscillators have the size of [-5D, 15D]×[-5D, 5D] and [-10D, 30D]×[-10D, 10D], respectively. For tandem oscillators, the initial center-to-center spacing between upstream and downstream cylinders has the value of $L = 1.5D$. The boundary conditions of flow domains and bluff oscillators are employed as follows: for inlet boundary, a Dirichlet boundary condition with $u_x = U_0$ and $u_y = 0$ for incoming flow is used; a constant pressure $P = 0$ and Neumann-type condition (i.e., $\partial u_x/\partial x = 0$, $\partial u_y/\partial x = 0$) are set at the outlet; no-slip boundary condition is applied in the cylinder surface.

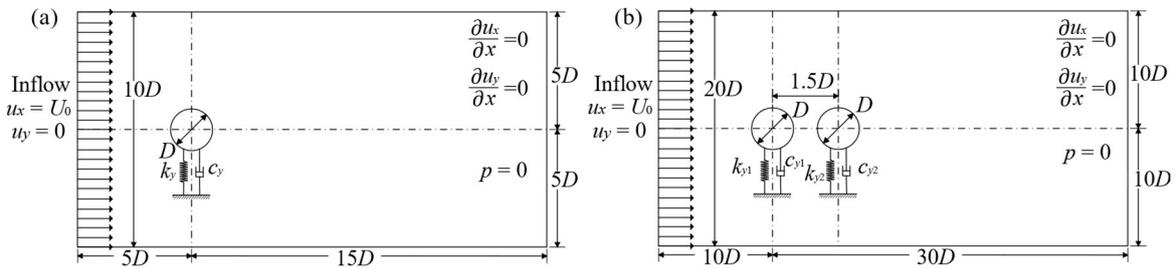

**Figure 5.** Schematic of computational domain and boundary conditions for VIV of cylinder: (a) an isolated cylinder (b) Two cylinders in tandem arrangement.

In this paper, the overset mesh technique is adopted to implement the movement of cylinder boundary in the flow field domain. Meanwhile, the flow field will evolve nonlinearly with the movement of the oscillator. Reasonably, the overset mesh technique belongs to the dynamic mesh technique and is suitable for rigid boundary motion problems, which has the advantages of high precision, high computation efficiency, and mesh non-distortion. As can be seen from Figure 6, the overset mesh consists of a component mesh enclosing the cylinder and a background mesh of the external flow field. Figure 6(a) and Figure 6(b) represent the concrete mesh for isolated cylinder and tandem cylinders, respectively. The solution steps of structural movement realized by the overset mesh are as follows: (1) Updating the position of the component mesh based on the cylinder's displacement drawn in Figure 4; (2) Identifying the overset mesh boundary by fluid solver; (3) Removing the part of the background mesh obscured by the component mesh; (4) Interpolating the variable information of the boundary elements in the background region to the boundary elements in the nested region; (5) Calculating the flow field.

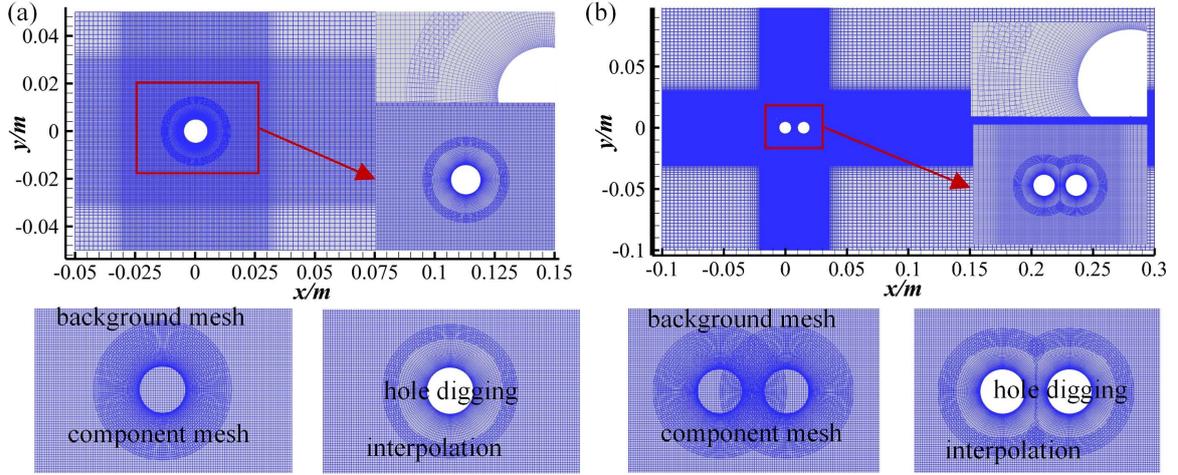

**Figure 6.** Numerical mesh around the cylinder: (a) an isolated cylinder (b) two cylinders in tandem.

The pressure/velocity fields, vibration displacement, and flow field force for both an isolated oscillator and two oscillators in tandem are simulated using the mentioned techniques. The accuracy and reliability of these acquired simulation results are verified through comparison with the existing numerical simulation results (Griffith et al., 2017; Zhao, 2022; Li et al., 2016), as listed in Table 1 and Table 2. As shown in Table 1 and Table 2, $y_{max}$, $f_y$ and $C_D$ represent the maximum amplitude, vertex shedding frequency, and drag coefficient, respectively. And corresponding subscripts, i.e., *mean* and *mse*, are the mean value and mean square error, respectively. For isolated cylinder with 2DOF, the various results have high consistency with that of Refs. (Zhao, 2022; Li et al., 2016). Under identical conditions, displacement and force information for the isolated oscillator with 1DOF are generated by restricting motion to the cross-flow direction. The motion in the direction of resistance will stimulate the displacement in the direction of lift, so the lower results of $y_{max}$ and $C_{L,mse}$ for 1DOF case in Table 1 (5$^{th}$ column) are reasonable. That is, the data generated for an isolated oscillator with 1DOF exhibit a high degree of accuracy. Similarly, high-precision results (in Table 2) are obtained for the upstream and downstream cylinders of the multi-vibration subsystem. Therefore, these credible VIV simulation results are regarded as the appropriate training dataset for the FSI model.

**Table 1.** Comparison of simulation indicators with existing data for an isolated cylinder with 1DOF.

| index | 2DOF Zhao, (2022) | 2DOF Li et al., (2016) | 2DOF-Present | 1DOF-Present |
|---|---|---|---|---|
| $y_{max}/D$ | 0.558 | 0.55 | 0.545 | 0.526 |
| $f_v/f_n$ | 0.980 | -- | 0.988 | 0.988 |
| $C_{D,mean}$ | 1.950 | 2.08 | 1.760 | 1.800 |
| $C_{L,mse}$ | 0.160 | 0.15 | 0.160 | 0.130 |

**Table 2.** Comparison of simulation indicators with existing data for two cylinders in tandem with 1DOF.

| | Upstream Cylinder | | | Downstream Cylinder | | |
|---|---|---|---|---|---|---|
| index | Zhao, (2022) | Griffith et al., (2017) | Present | Zhao, (2022) | Griffith et al., (2017) | Present |
| $y_{max}/D$ | 0.67 | 0.72 | 0.72 | 0.63 | 0.65 | 0.71 |
| $f_v/f_n$ | 0.86 | -- | 0.86 | 0.86 | -- | 0.86 |
| $C_{D,mse}$ | 1.67 | 1.79 | 1.74 | 1.02 | 1.12 | 1.08 |
| $C_{L,mse}$ | 0.79 | 0.84 | 0.79 | 0.73 | 0.78 | 0.73 |

*3.3. Data processing and verification*

To unify the input and output of the proposed FSI network model in the form of two-dimensional matrix pixel points with fixed positions, in this paper, the nearest neighbor interpolation method is adopted to process the irregularly distributed and time-varying flow field points outputted by CFD solver as the size of $N_x \times N_y = 256 \times 256$, as shown in Figure 7. Here, the flow field dimensions of the isolated oscillator and the double oscillators used for modeling are selected in the ranges of [-2.3$D$, 6.7$D$] × [-2.33$D$, 2.33$D$] and [-1.0$D$, 2.84$D$] × [-1.92$D$, 1.92$D$], respectively, so as to better analyze the intense flow field evolution around the cylinder. As a matter of fact, the vicinity of the cylinder contains a higher concentration of complex flow field characteristics, necessitating a finer interpolation grid to accurately represent these intricate physical features. Correspondingly, the minimum size of the interpolation mesh for the isolated oscillator and double oscillators is set as 0.013$D$ and 0.015$D$, respectively. We compare the pressure field and two velocity fields processed by the nearest neighbor interpolation method with respect to that of the full-order grid outputted directly from the CFD solver. Following Figure 8, the two methods have a highly similar flow field distribution especially the contour trend, meaning that the chosen interpolation method is appropriate for the fitting of the original full-order flow field.

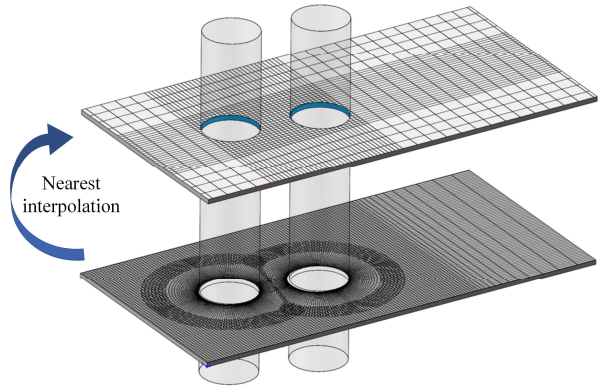

**Figure 7.** Data processing realized by the nearest interpolation for the grid information of two cylinders in tandem arrangement.

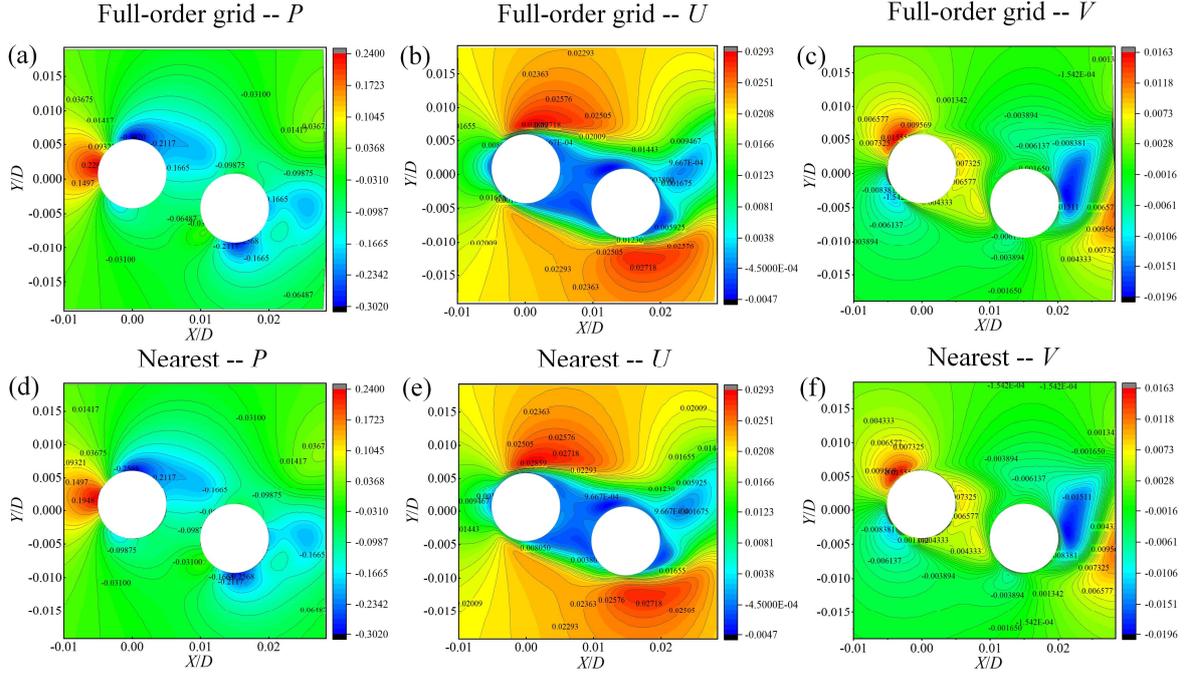

**Figure 8.** Comparison of pressure field and velocity fields processed by the nearest neighbor interpolation method with respect to that of full-order grid outputted directly from CFD solver at time step $tU_0/D=60$: (a) full-order grid for pressure field (b) full-order grid for velocity field in X-direction (c) full-order grid for velocity field in Y-direction (d) interpolation method for pressure field (e) interpolation method for velocity field in X-direction (f) interpolation method for velocity field in Y-direction.

Besides, to further confirm the high fidelity of the processed fields around the oscillators, it is essential to compare the cylindrical lift force calculated by the treated flow field with directly outputted from the CFD solver. Meanwhile, the above operation is indispensable in terms of ensuring the accuracy of the FSI model in the interaction of force information, as constructed in Figure 1(b). Different from previous study (Han et al., 2022) in force calculation, apart from integrating the discrete pressure points on the cylindrical surface, the wall friction, as one of the essential force components, solved by wall shear stress is also introduced here. Therefore, a general approximate circular arc method based on discrete points is presented in this paper, as shown in Algorithm 1. The schematic diagram of the discrete integral is depicted in Figure 9. The results shown in Figure 10 - Figure 11 indicate that the proposed arc method combined with pressure field and wall shear stress has excellent performance and is suitable for force calculation. Additionally, in Figure 10 and Figure 11, for the time steps $tU_0/D$ from initial to stable vibration, the fitting degree between the force calculated with only the pressure field and the labeled value of numerical simulation is not optimistic, which intuitively proves the significance of wall shear stress in this paper, even for laminar flow.

Algorithm 1. Force calculation based on discrete points.

> **Input**: Two-dimensional pressure field $P$; center position $(x_i, y_i)$ of the oscillator; discrete wall shear stress ($\bar{\tau}$)
> **Output**: Lift force ($F_L$) & Drag force ($F_D$)
> **Process A: Forces caused by pressure field**
> 1. In Figure 9(a), find the flow field points of the pressure field closest to the cylindrical outer surface, including the pressure value and corresponding coordinate for each point: $\bar{P}$, $\bar{X}$, $\bar{Y}$

2. Take the flow field points closest to the *X*-axis in the second quadrant as the starting point, arrange the points in step 1 in a clockwise direction to generate a new sequence: $\overline{P}=\{p_1\ p_2\ldots p_i\ldots p_n\}$, $\overline{X}=\{x_1\ x_2\ldots x_i\ldots x_n\}$, $\overline{Y}=\{y_1\ y_2\ldots y_i\ldots y_n\}$

3. According to $\overline{X}$ and $\overline{Y}$, the angle of the line between the flow field point and the center of the circle relative to the negative half axis of the *X*-axis is obtained: $\overline{\theta}=\{\theta_1\ \theta_2\ldots\theta_i\ldots\theta_n\}$, as shown in Figure 9(a).

4. Combined with the schematic diagram of discrete force solution in Figure 9(b), the pressure at each point (i.e., $F_{Pi}=\sqrt{F_{LPi}^2+F_{DPi}^2}$) is calculated and summed to obtain the lift force $F_{LP}$ and drag force $F_{DP}$.

**Process B: Forces caused by wall shear stress**

1. Different from the discrete point distribution of the pressure field, the points of the wall shear stress are directly output by the CFD solver and arranged on the cylindrical surface with equal spacing: $\overline{\tau}=\{\tau_1\ \tau_2\ldots\tau_j\ldots\tau_k\}$, as drawn in Figure 9(c). Therefore, the value of $\overline{\varphi}=\{\varphi_1\ \varphi_2\ldots\varphi_i\ldots\varphi_k\}$ can be determined.

2. The lift force $F_{L\tau}$ and drag force $F_{D\tau}$ summed by the discrete wall friction (i.e., $F_{\tau j}=\sqrt{F_{L\tau j}^2+F_{D\tau j}^2}$, as shown in Figure 9(d)) can be solved.

* The $F_L$ is composed of two elements: $F_{LP}$, $F_{L\tau}$, as expressed in Equation (37) and Equation (38).

* In this paper, only the solution of lift is considered.

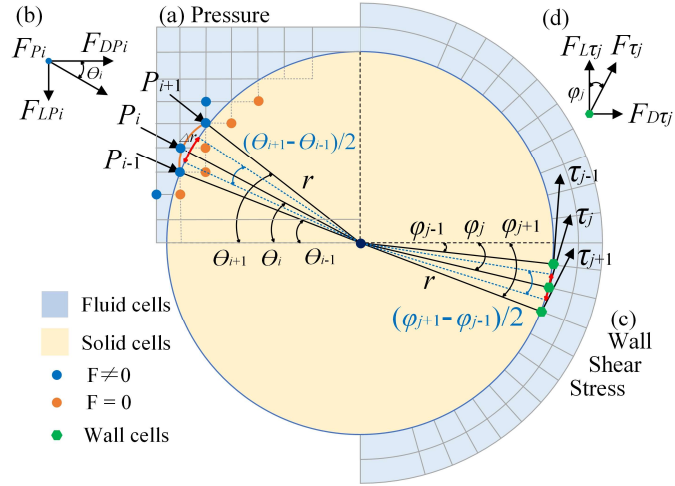

**Figure 9.** Schematic diagram of the discrete integral of lift and drag for oscillating cylinder--Arc Method: (a) integration of the discrete pressure points (b) pressure decomposition (c) integration of the discrete stress points (d) wall shear stress decomposition.

$$\begin{cases} F_{LP}=\sum_{i=1}^{n} F_{LP_i}=\sum_{i=1}^{n} F_{P_i}\times\sin\theta_i\approx\sum_{i=1}^{n} 0.5\times P_i\times r\times(\theta_{i+1}-\theta_{i-1})\times\sin\theta_i \\ F_{L\tau}=\sum_{j=1}^{m} F_{L\tau_j}=\sum_{j=1}^{m} F_{\tau_j}\times\cos\varphi_j\approx\sum_{j=1}^{m} 0.5\times\tau_j\times r\times(\varphi_{j+1}-\varphi_{j-1})\times\cos\varphi_j \end{cases} \quad (37)$$

$$F_L=F_{LP}+F_{L\tau}. \quad (38)$$

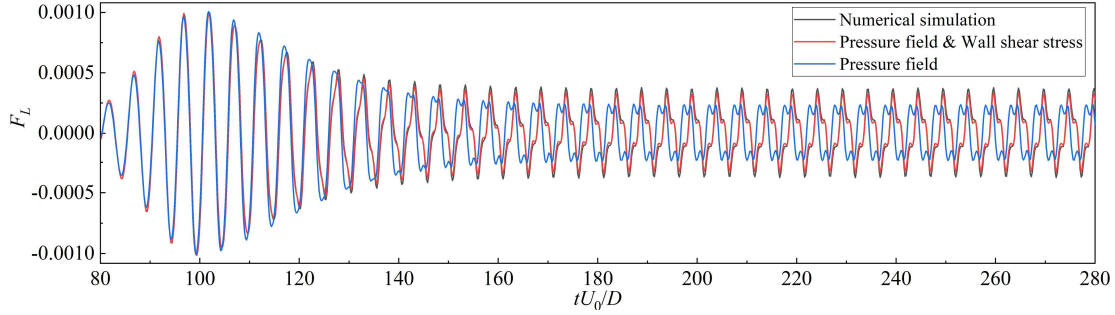

**Figure 10.** Comparison of lift force of an isolated cylinder between the numerical simulation and data processing ('red line': integrating the surface pressure and wall shear stress; 'blue line': integrating only the surface pressure).

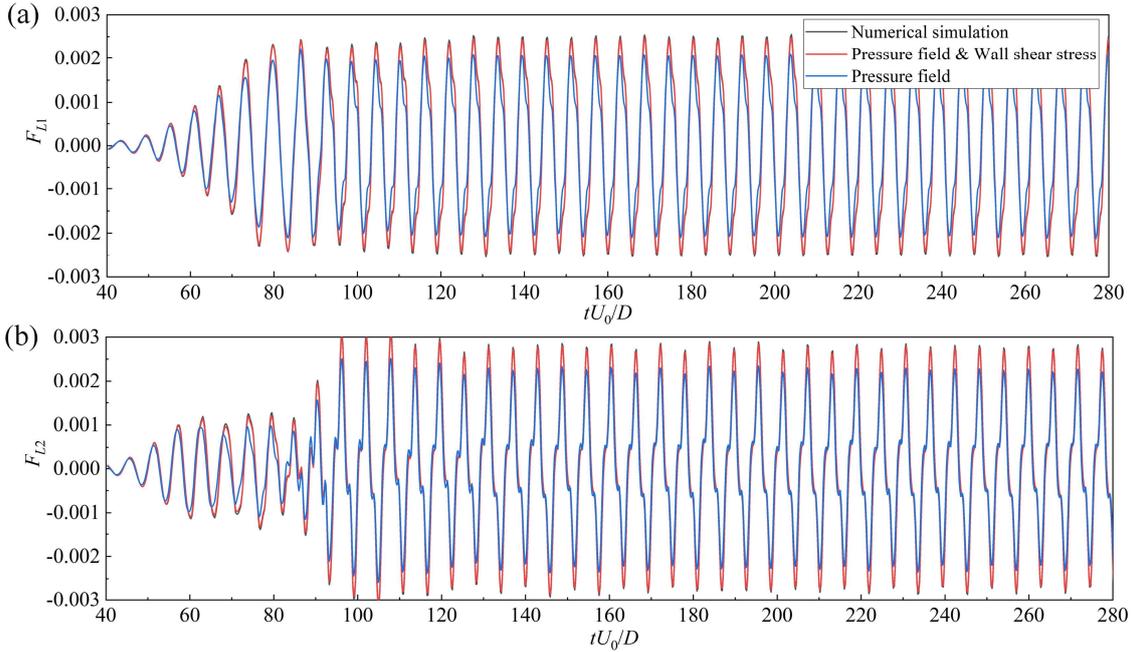

**Figure 11.** Comparison of lift force between the numerical simulation and data processing ('red line': integrating the surface pressure and wall shear stress; 'blue line': integrating only the surface pressure): (a) upstream cylinder (b) downstream cylinder.

## 4. Results and discussion

In this section, we explore and verify the modeling accuracy and reliability of the proposed USFNO-FConvLSTM model and USFNO-based hybrid neural network model in predicting VIV evolution of an isolated cylinder or two cylinders in tandem arrangement under chosen implementation details and evaluation metrics.

*4.1. Implementation details and evaluation metrics*

For each modeling, thousands of consecutive time snapshots (samples) with equal spacing of 0.1s are split into two phases for training and testing. We construct the models for 200 epochs by training 800 samples. The sample size applicable to the prediction will vary depending on the specific model. The RMSProp optimizer with a hyperparameter of 0.9 and initial learning rate of 0.0005 is utilized to optimize network parameters. All experiments are implemented on Pytorch architecture and trained on an NVIDIA Tesla V100S GPU with 32GB VRAM and 100 GB RAM.

To evaluate the modeling reliability comprehensively, we employ three evaluation metrics including mean absolute error (MAE), mean relative error (MRE) based on norm, and root mean square error (RMSE). Among them, MAE is selected to assess the deviation extent of label data $\hat{U}_i$ compared with the prediction value $U_i$, and is defined as:

$$\text{MAE} = \frac{1}{N}\sum_{i=1}^{N}|\hat{U}_i - U_i|. \tag{39}$$

MRE can intuitively reflect the prediction accuracy in the proportion form. To avoid meaningless calculations caused by cylinder filling zero value as the denominator, this paper adopts a modification employing the 1-norm to the conventional MRE and denotes it as:

$$\text{MRE}(\%) = \sum_{i=1}^{N}|\hat{U}_i - U_i| / N \sum_{i=1}^{N}|\hat{U}_i| \times 100\%. \tag{40}$$

RMSE is another common evaluation metric for the regression task. Compared to MAE, RMSE is sensitive to large errors and is greatly influenced by outliers. The formula is as follows:

$$\text{RMSE} = \sqrt{\frac{1}{N}\sum_{i=1}^{N}(\hat{U}_i - U_i)^2}. \tag{41}$$

*4.2. Network parameter determined of USFNO-FConvLSTM*

As described in Section 2, the innovative USFNO-FConvLSTM model allows sequence-to-point training and prediction by fusing historical time sequence features in the FConvLSTM module. Consequently, it is essential to analyze the effect of the fusion time steps on the modeling accuracy and explore the superiority of multiple moving boundary systems that the FConvLSTM architecture in contrast to the conventional ConvLSTM framework. It is noteworthy that the process of determining the model parameters of USFNO-FConvLSTM is exclusive of the structural motion equation and its feedback. Taking the VIV of two cylinders as the research object, the prediction dataset comprises 200 consecutive time snapshots, corresponding to the time step $tU_0/D$ from 201 to 241.

*4.2.1. Fusion time step*

Here, an isolated FConvLSTM layer is chosen to avoid excessive modeling time for the training process. Six kinds of exponentially increasing fusion time steps, i.e., $N = 1, 2, 4, 8, 16$, and $32$, are applied in this paper. Figure 12 illustrates the average MAE, MRE, and RMSE with the increasement of time steps $tU_0/D$ in the prediction process for pressure and velocity fields. Focusing on pressure field prediction, as depicted in Figure 12(a,b,c), when $tU_0/D$ is higher than 208, increasing the fusion steps from 1 to 32 will induce all three metrics to increase first and then decrease. The minimum error points occur at $N = 16$, with the maximum at $N = 1$, reflecting variable performance with step increments. A similar phenomenon can be observed for the velocity fields in both the *X*- and *Y*-direction in Figure 12(d,e,f,g,h,i). Corresponding minimum and maximum errors will appear when $N = 16$ and $N = 4$, respectively. Therefore, the fusion of historical time-step information within the appropriate range is more conducive to accurate modeling than the single time-step iteration imitating the traditional CFD method. Summarily, the optimal fusion time step for the two-cylinder system is determined as $N = 16$. The above reasonable phenomena can be explained by the VIV period. Concretely, the VIV system with two-cylinder has the natural frequency of $f_{ny} = 0.40192$Hz, corresponding to the vibration period of $T = 1/f_{ny} \approx 2.5$s. Accordingly, each flow cycle contains approximately 25 snapshots spaced at 0.1s intervals. The fusion time step of $N = 16$ is closer to 25 compared to steps $N = 1\text{-}8$, indicating that its fused flow field features are more comprehensive and effective. Nevertheless, the deep model with $N = 32$ consistently performs worse capability compared with $N = 16$ because the premature time information causes noise interference to the prediction.

In Figure 12, as $tU_0/D$ increases, the average MAE initially decreases followed by a stable fluctuation with an insignificant decline in prediction accuracy, which is attributed to the excellent forecasting capability of the established USFNO-FConvLSTM model and the stable sensitivity of MAE to error outliers at cylindrical boundaries. Simultaneously, MRE and RMSE both display an escalating trend with the increasing of $tU_0/D$. This phenomenon is associated with the error accumulation effect that the prediction of future fluid features relies on historical information. It is also relevant to the principle that MRE and RMSE are more sensitive to error fluctuations and the detection of maximum error than MAE. In conclusion, $N = 16$ is chosen as the fusion time step for further study in this paper.

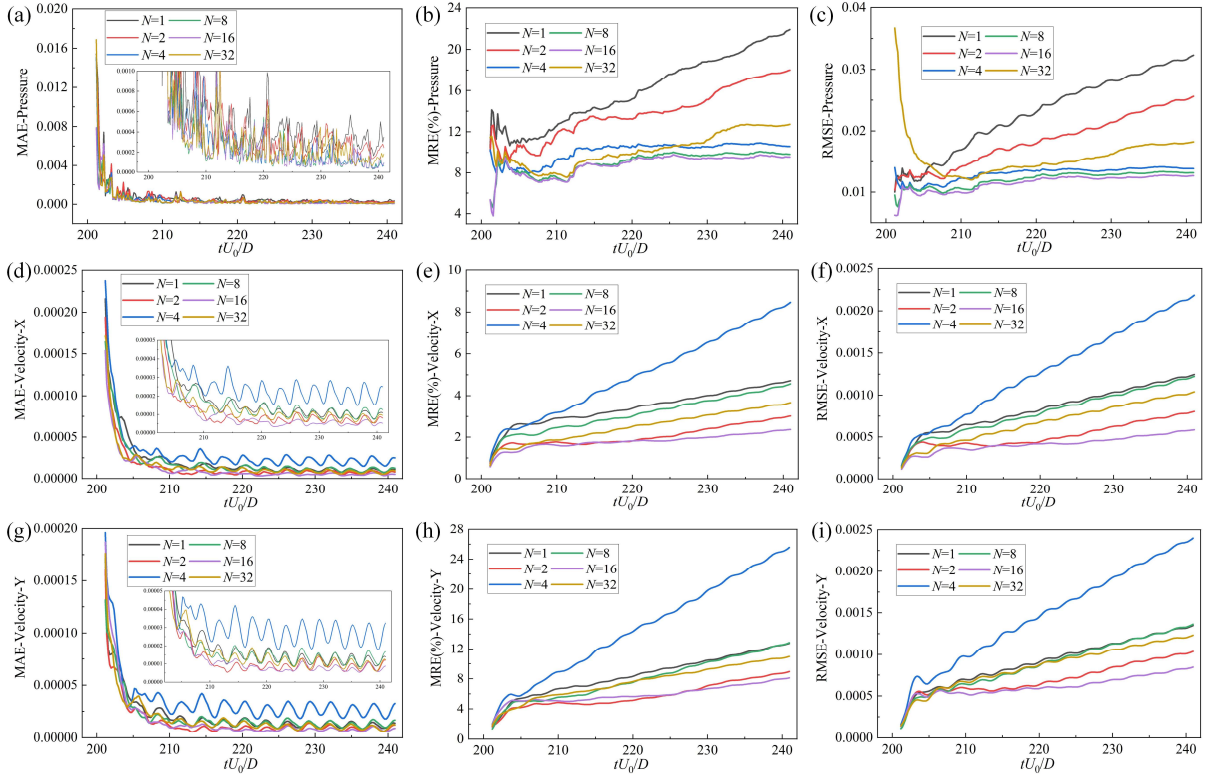

**Figure 12.** Three evaluation metrics of flow fields predicted by the USFNO-FConvLSTM model with different numbers of fusion time steps for two cylinders in tandem: (a) MAE-Pressure (b) MRE-Pressure (c) RMSE-Pressure (d) MAE-Velocity-X (e) MRE-Velocity-X (f) RMSE-Velocity-X (g) MAE-Velocity-Y (h) MRE-Velocity-Y (i) RMSE-Velocity-Y.

*4.2.2. FConvLSTM vs. ConvLSTM vs. FNO*

To reflect the rationality and superiority of the FConvLSTM fused multiple time steps ($N = 16$), the existing ConvLSTM architecture realized by successively entering the multiple time steps information in sequence [35] is combined with the proposed USFNO framework, forming the USFNO-ConvLSTM model. Subsequently, a comparative analysis is conducted between this model and the USFNO-FConvLSTM, emphasizing the distinctions between these two and the advantages of the fusion strategy. Emphatically, the above two novel frameworks set the unified network architecture except for the time processing module. Moreover, to demonstrate the superiority of USFNO integrated with the time series network, a similar sequence-to-point prediction model is constructed utilizing the original FNO as presented by Li et al. (2021), which is comprised solely of stacked Fourier layers. All models are trained on the identical implementation specifics mentioned as Section 4.1, with a comprehensive intercomparison through the analysis of training loss evolution over epochs and prediction error evolution against $tU_0/D$, as shown in Figure 13. Figure 13(a) indicates that, for the training process, in contrast with the other two models, the fastest convergence speed and lowest training error are accomplished by the USFNO-FConvLSTM model. Besides that, the USFNO-FConvLSTM model has the best performance with respect to modeling accuracy and mitigating effect on error accumulation during the prediction phase, as evidenced by the gentle trend of its average error curves, as shown in Figure 13(b,c,d). This excellent performance is particularly notable in comparison to the USFNO-ConvLSTM in the initial prediction and FNO in the terminal forecast, as shown in Figure 13(c,d). The corresponding visualized predictive pressure fields are illustrated in Figure 14. The USFNO-FConvLSTM model demonstrates a remarkable capability in accurately capturing the moving boundaries of the cylinders, which correspond to regions of significant fluid flow gradient variations ($2^{th}$ and $5^{th}$ columns). Moreover, it is capable of fitting the flow field characterized by smooth isobars, aligning more in conformity with physical mechanism, especially in areas featured by high-pressure and low-pressure. As depicted in Figure 14 ($3^{rd}$ and $6^{th}$ columns), the ConvLSTM-based model shows relatively inferior predictive performance in dynamic boundaries and pressure field distribution compared with FConvLSTM-based model. Nevertheless, the prediction effects of the FNO-based model for flow field and dynamic boundary are evidently not within the acceptable range, that is, showing deviation boundaries and zigzag isobars ($4^{th}$ and $6^{th}$ columns).

To sum up, the USFNO-FConvLSTM model has an excellent forecast ability than that of the USFNO-ConvLSTM model on account of directly fusing continuous historical time information. Concretely, as $tU_0/D$=241, the MAE, MRE, and RMSE of the former are reduced by 50.8%, 26.5%, and 34.5%, respectively, compared with the latter. Interestingly, although the USFNO-ConvLSTM model shows the worst capability relative to other models at the initial prediction, it has a more stable and lower error than that of the FNO model in the subsequent prediction phase. The FNO model here generates the highest error in comparison to the other two models due to its finite Fourier basis and the inherent regularization effect, which further demonstrates the validity of the novel USFNO framework and FConvLSTM.

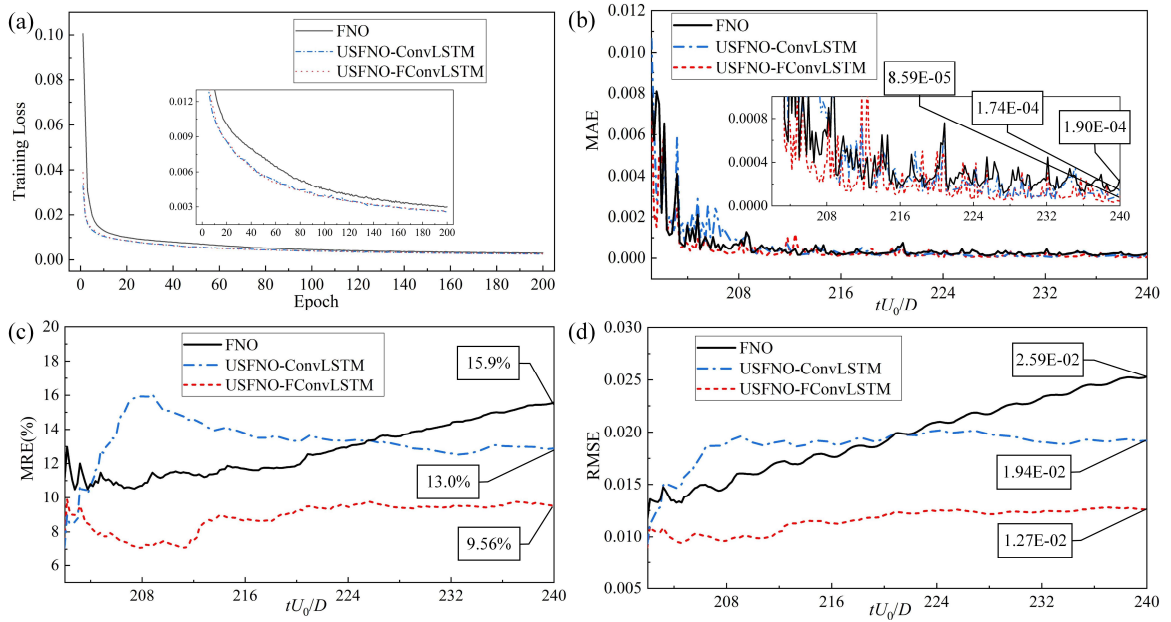

**Figure 13.** Comparison of error distribution of pressure field set by FNO, USFNO-ConvLSTM, and USFNO-FConvLSTM models: (a) training loss vs. epoch (b) MAE vs. $tU_0/D$ (c) MRE vs. $tU_0/D$ (d) RMSE vs. $tU_0/D$.

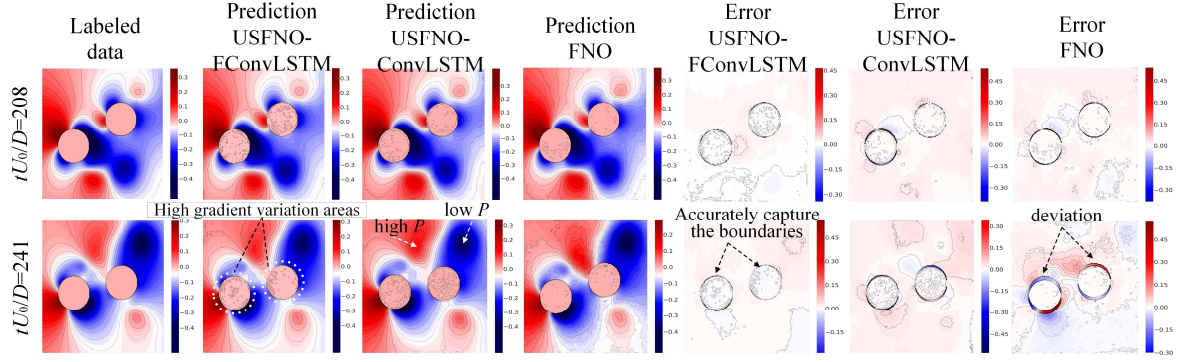

**Figure 14.** Qualitative comparison of the spatial-temporal predicted pressure field and its error field for USFNO-FConvLSTM, USFNO-ConvLSTM, and FNO models. The plots presented at dimensionless time instants $tU_0/D$=208 and 241 for two tandem-cylinders.

*4.3. The comparison between USFNO-FConvLSTM, Unet-FConvLSTM, and CNN-FConvLSTM*

As an essential component in the USFNO-based hybrid neural network model for precise modeling of flow field, the novel USFNO-FConvLSTM model implemented for parameter determination in Section 4.2 is proven to have the excellent accuracy and reliability of spatial-temporal flow field evolution. Besides, two newly FConvLSTM-based models respectively composed with CNN and Unet, i.e., CNN-FConvLSTM (Figure A.1 in Appendix) and Unet-FConvLSTM (Figure A.2 in Appendix), are presented here to make a comparison with the USFNO-FConvLSTM model. A series of comparison curves of evaluation metrics for pressure and velocity fields at various learning rates under three different models are drawn in Figure 15. The corresponding error ratios between the models are depicted in Figure 16. The evolution tendency of predicted average errors with increasement of $tU_0/D$ between the USFNO-FConvLSTM and CNN-FConvLSTM is shown in Figure 17. In addition, Figure 18 - Figure 20 qualitatively illustrate and compare the predicting ability for spatial-temporal flow field evolution using the above three presented models. Corresponding visualization results contain the variables of $tU_0/D$ = 207, 216, 227, 236, and 241. Being similarly to Section 4.2, VIV of two oscillators in tandem is taken as the modeling object, and continuous time series here, which accounted for 20% of the total number of 1000, are used for modeling generalization evaluation.

**USFNO-FConvLSTM vs. CNN-FConvLSTM**: From a quantitative perspective, the USFNO-FConvLSTM model obtains the lower prediction errors at all the chosen learning rates in comparison to the CNN-FConvLSTM model, as shown in Figure 15. Taking the pressure field for instance, as $lr$ = 0.0005, the average MAE, MRE, and RMSE of the former are $8.58\times10^{-5}$, 9.56%, and $1.27\times10^{-2}$, respectively. These errors are significantly lower than that of the latter model, showing reductions of 45.90%, 38.80%, and 45.13% respectively. The more intuitive error ratio $E$ defined in Equation (42) is shown in Figure 16.

$$E_B^A = (A-B)/B \times 100\%. \tag{42}$$

where $A$ denotes the prediction error of the USFNO-FConvLSTM model; and $B$ represents the value for one of the other two models. As illustrated in Figure 16, a negative value for $E_B^A$ signifies that the USFNO-FConvLSTM model outperforms the others, whereas positive value indicates the more superiority for other two models. For instance, when $B$ denotes CNN-based model in Figure 16, apart from the MAE ratio as $lr$ = 0.001, all the values of $E_B^A$ are lower than zero, meaning that the USFNO-based model has superior performance. As drawn in Figure 17, for CNN-based model, with the increase of $tU_0/D$ during prediction process, the average MAE and MRE of velocity fields in the $X$- and $Y$-directions have both an overall upward trend and eventually reach up to the value that cannot reasonably predict the flow field evolution. This phenomenon is consistent with the spatial-temporal velocity fields visualized in Figure 19 and Figure 20. Concretely, in Figure 19 and Figure 20, the CNN-FConvLSTM model fails to adequately fit the upstream cylindrical wake field featuring high non-linearity (4$^{th}$ and 7$^{th}$ columns) and even generates incapacitation to capture the downstream cylinder and its dynamic boundary with the advance of prediction. In addition, the CNN-based model exhibits significant prediction errors in the flow field, as well as deviations between the cylindrical position and its label value (7$^{th}$ column). Contrary to the forecasting variable tendency of velocity fields for the CNN-FConvLSTM model, a series of steady error curves (represented by dashed line) with almost no upward trend can be obtained by the USFNO-FConvLSTM model, as shown in Figure 17. Correspondingly, a meaningful phenomenon can be observed in Figure 18 - Figure 20 that the USFNO-FConvLSTM model demonstrates its capability to accurately predict the oscillatory boundaries of tandem bicylinders with phase differences, along with the distribution of pressure and velocity fields around the oscillators. The complex interactions between the cylinders are visualized through the flow fields evolution in the gaps, and the precise predictions realized by the USFNO-based model in this region fundamentally demonstrate its capability for accurately predicting these interactions—an ability that is not possessed by the CNN-based model. The above comparison indicates the excellent prediction ability of USFNO-FConvLSTM architecture for nonlinear complex moving boundary and the superiority of USFNO module over CNN. As expected, the subpar performance of the CNN-based model in spatiotemporal prediction problems for multi-cylinders can be attributed to its local regularization, leading to the issues to predict the complex and global physical system accurately.

**USFNO-FConvLSTM vs. Unet-FConvLSTM**: To reflect the operation of coupling dimensionality-reduced information with dimensionality-increased features achieved by U-shaped structure to enhance the prediction ability of CNN-based model, this paper configures the network frameworks of CNN-FConvLSTM and Unet-FConvLSTM to be consistent, except for the U-shaped part. It can be evidently observed that the Unet-based model has the preferable modeling performance than that of the CNN-based one, especially for two velocity fields, as depicted in Figure 15 and Figure 18 - Figure 20. It signifies that the Unet-framework applied in this paper is effective for predicting complex interference between two tandem-cylinders. Nevertheless, the Unet-FConvLSTM model is still flawed in predicting pressure field that contains a richer array of physical features. More concretely, the time-lag phenomenon, i.e., a non-negligible phase difference between the predicted position and label position, can be found in the error field (6$^{th}$ column of Figure 18). The USFNO-FConvLSTM model avoids this delayed effect, primarily due to its foundation on FNO principles that integrate the linear global integral operator with the nonlinear local activation function. The superior capability of the USFNO-FConvLSTM model in precisely capturing the dynamic boundaries is also attributed to the FNO's excellent learning ability for high-frequency edge information (i.e., moving cylindrical edge). Beyond that, with the exception of selected individual learning rates, modeling errors of the USFNO-FConvLSTM consistently remain lower than those of the Unet-FConvLSTM across nearly all cases, as exhibited in Figure 15 and Figure 16.

To sum up, it can be concluded that the USFNO-FConvLSTM architecture can extract and combine the preponderances of Fourier layer and U-shaped network structure to complete the accurate fitting of highly-nonlinear global flow field and local dynamic boundary for spatial-temporal VIV evolution of multiple moving objects with complex interaction between each other. Furthermore, for solving the flow field evolution over the same time period, the USFNO-FConvLSTM model exhibits significantly greater predictive efficiency in comparison to conventional FSI numerical simulations. For instance, the former can be predicted 4020 times faster than the latter. It should be emphasized that it is indispensable to analyze and verify the USFNO-FConvLSTM's modeling ability to capture the dynamic boundary because it is directly related to whether the cylindrical force with the small deviation can be obtained. In other words, it is a pivotal prerequisite for the formation of a reliable FSI network model. Based on the above exploration, we conducted a study on the USFNO-based hybrid neural network, as described in the following section.

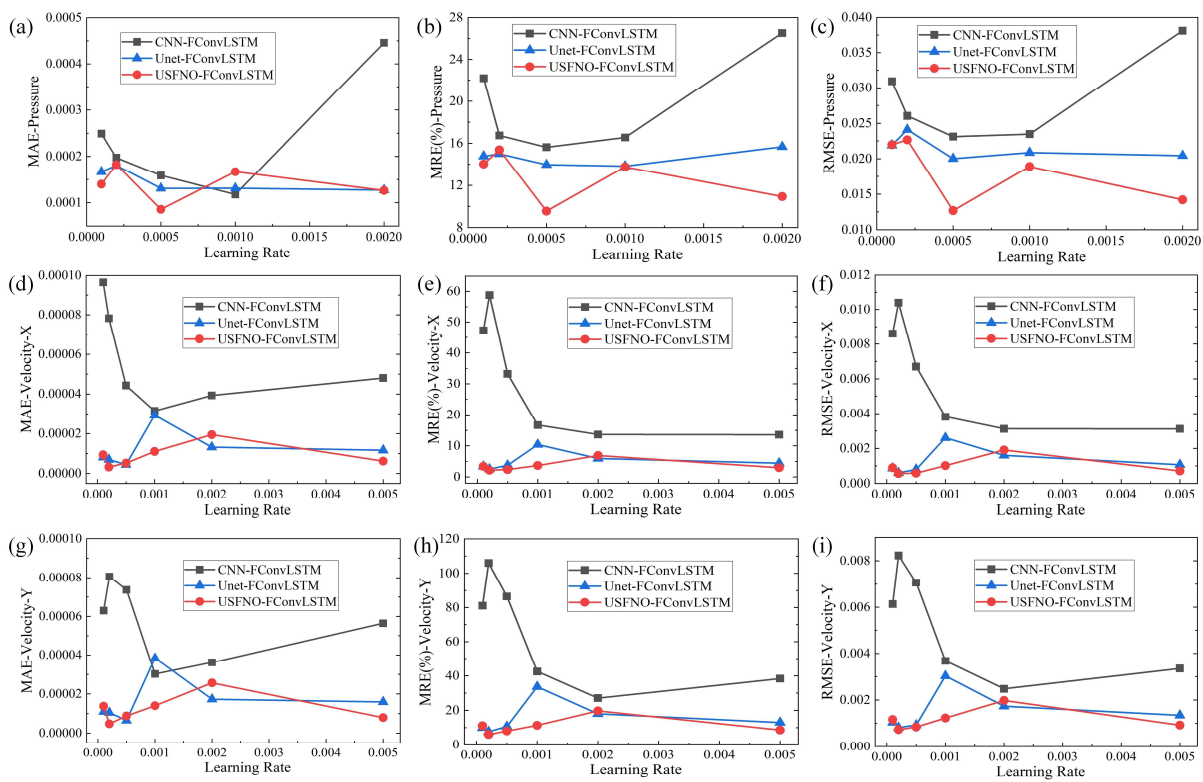

**Figure 15.** Three evaluation metrics of flow fields predicted by the USFNO-FConvLSTM, Unet-FConvLSTM, and CNN-FConvLSTM with different learning rates for two cylinders in tandem: (a) MAE-Pressure (b) MRE-Pressure (c) RMSE-Pressure (d) MAE-Velocity-X (e) MRE-Velocity-X (f) RMSE-Velocity-X (g) MAE-Velocity-Y (h) MRE-Velocity-Y (i) RMSE-Velocity-Y.

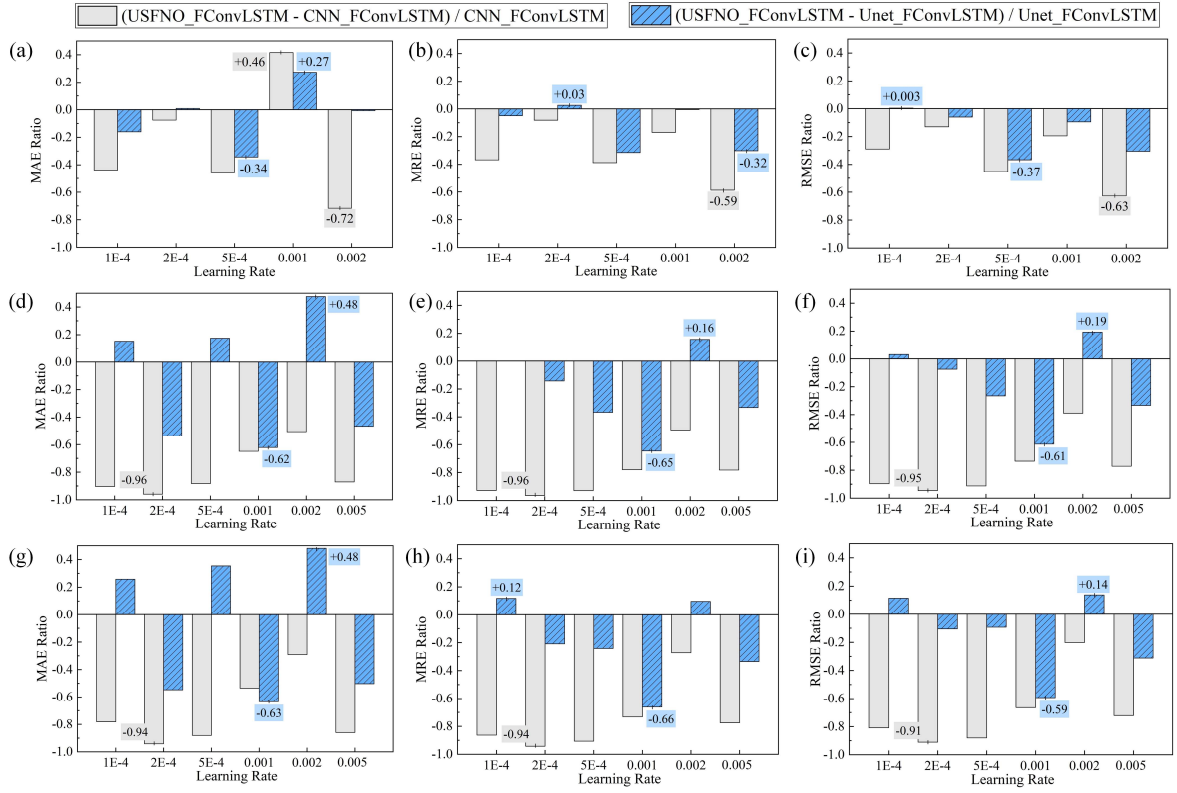

**Figure 16.** Three error ratios of flow fields compared between the USFNO-FConvLSTM, Unet-FConvLSTM, and CNN-FConvLSTM with different learning rates for two cylinders in tandem: (a) MAE-Pressure (b) MRE-Pressure (c) RMSE-Pressure (d) MAE-Velocity-X (e) MRE-Velocity-X (f) RMSE-Velocity-X (g) MAE-Velocity-Y (h) MRE-Velocity-Y (i) RMSE-Velocity-Y.

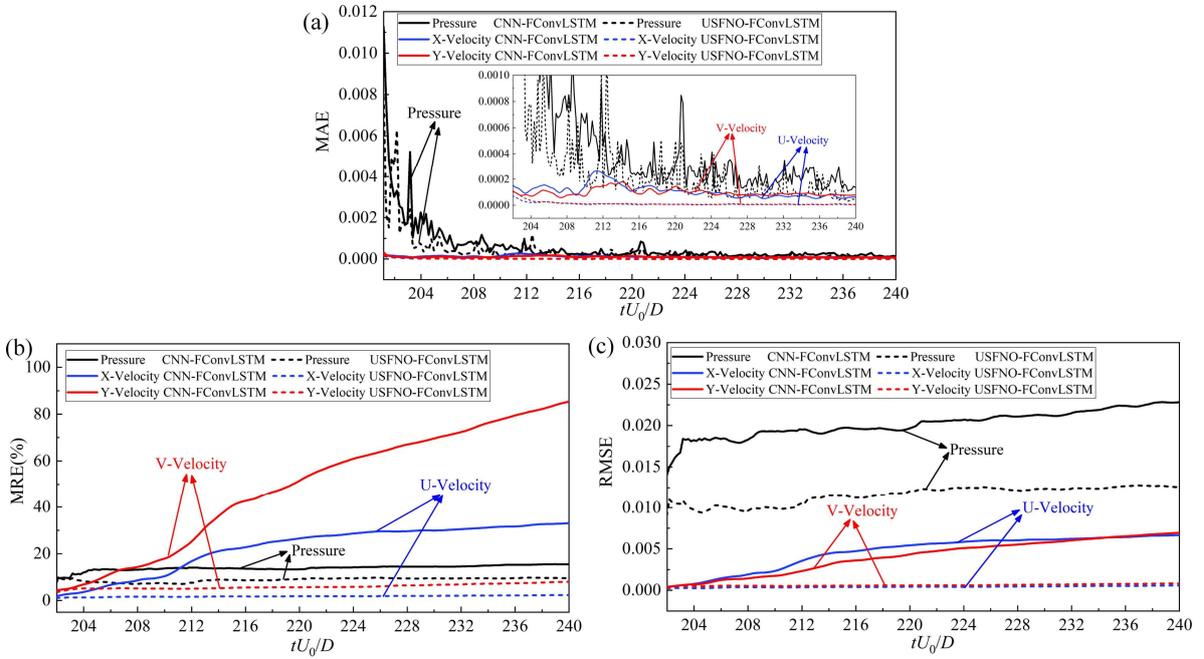

**Figure 17.** Comparison of average error distribution of pressure and velocity fields predicted by the proposed USFNO-FConvLSTM and CNN-FConvLSTM: (a) MAE vs. $tU_0/D$ (b) MRE vs. $tU_0/D$ (c) RMSE vs. $tU_0/D$.

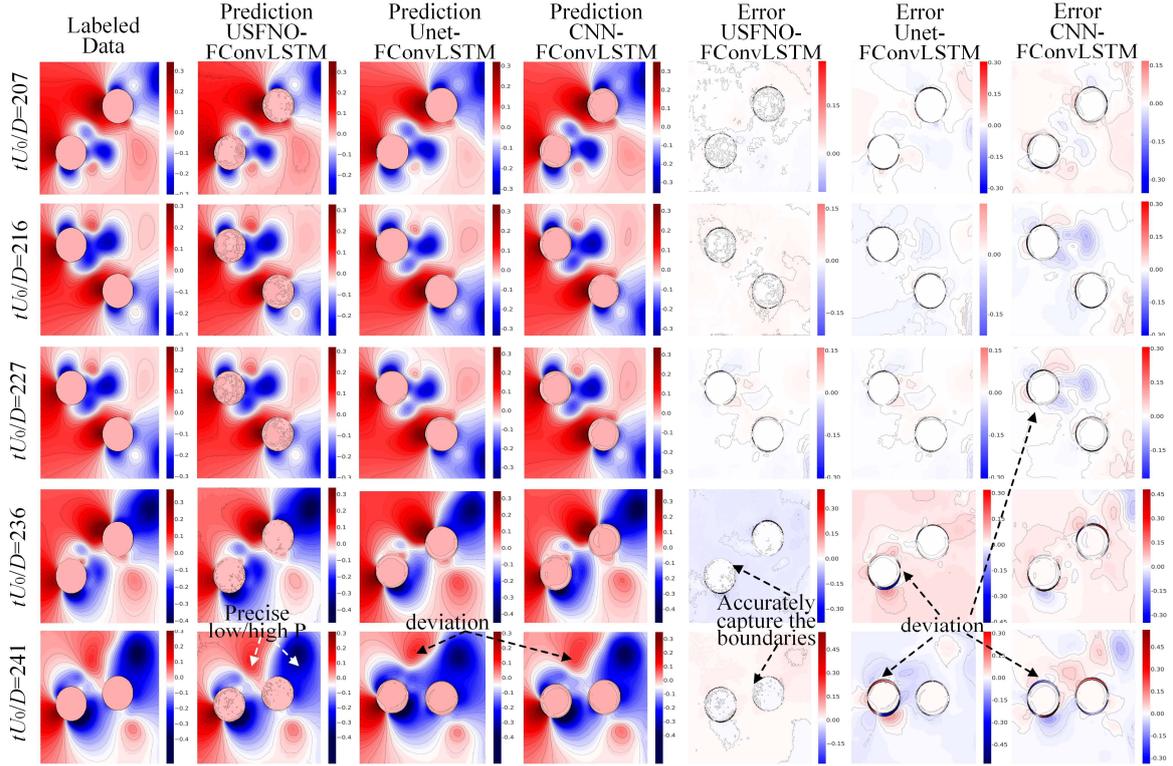

**Figure 18.** Qualitative comparison of the spatial-temporal predicted pressure field and its error field for USFNO-FConvLSTM, Unet-FConvLSTM, and CNN-FConvLSTM models. The plots presented at dimensionless time instants $tU_0/D$=207, 216, 227 and 241 for two tandem-cylinders.

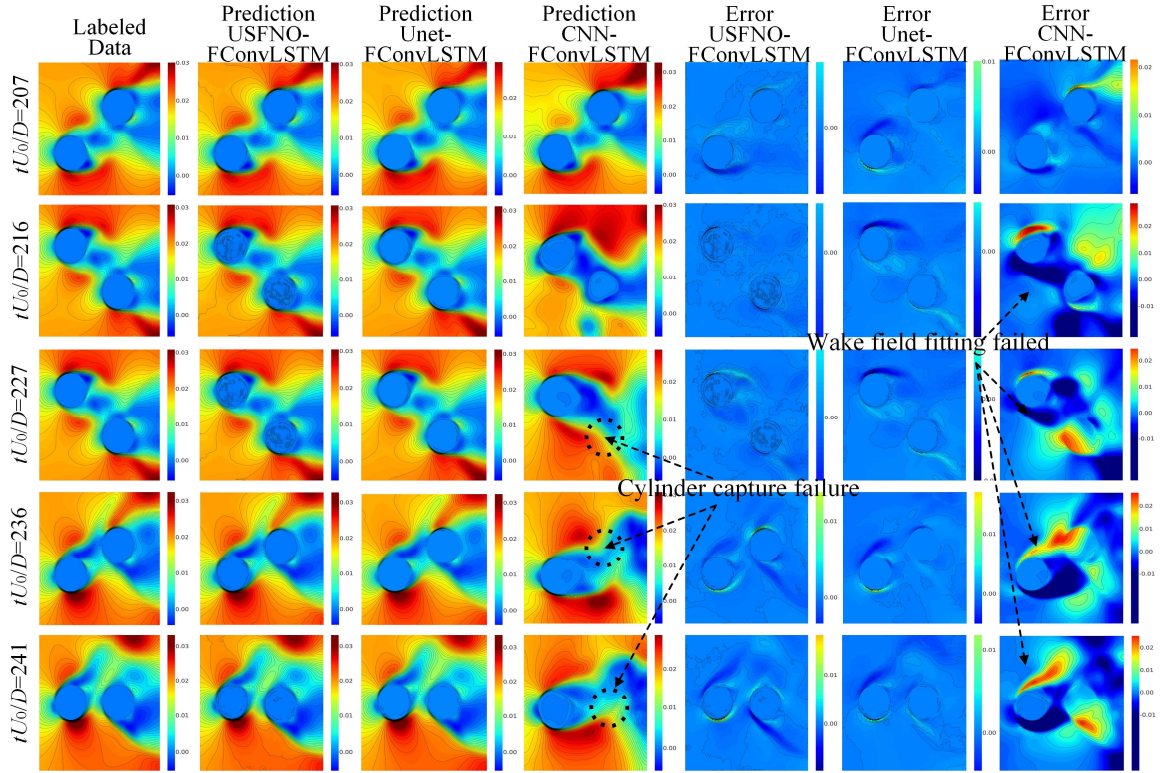

**Figure 19.** Qualitative comparison of the spatial-temporal predicted velocity field (X-direction) and its error field for USFNO-FConvLSTM, Unet-FConvLSTM, and CNN-FConvLSTM models. The plots presented at dimensionless time instants $tU_0/D$=207, 216, 227 and 241 for two tandem-cylinders. The black dotted circle indicates the cylindrical location of the labeled field.

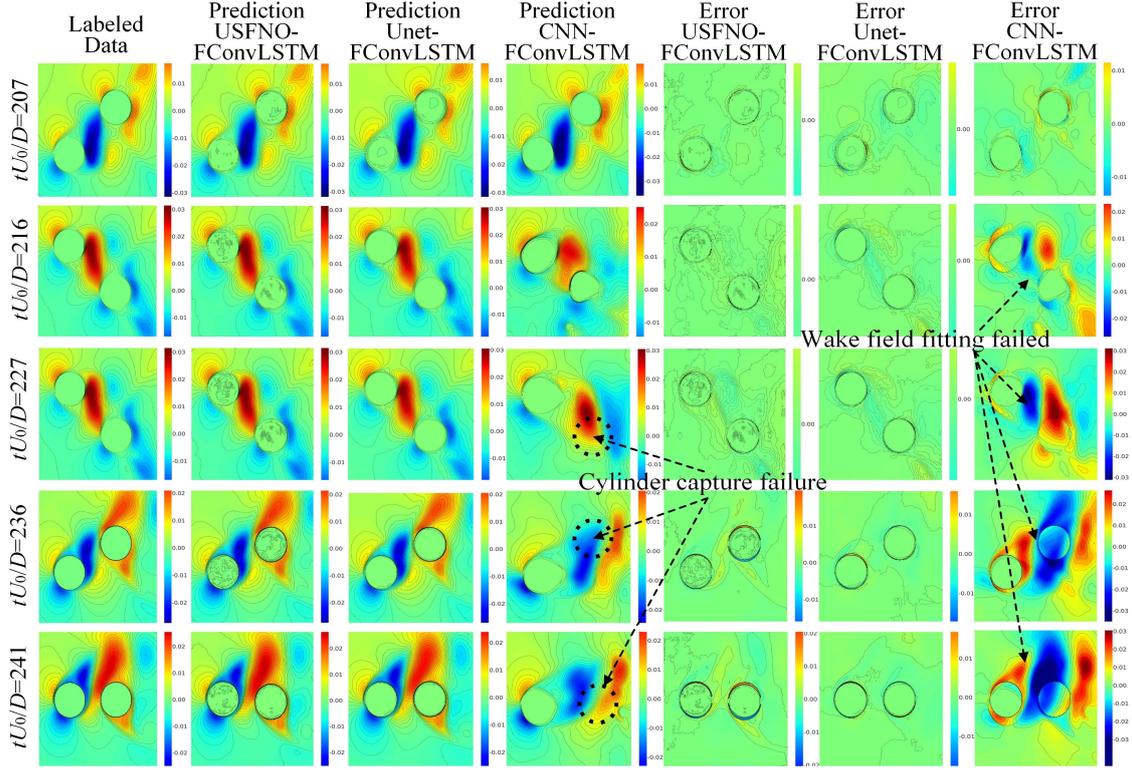

**Figure 20.** Qualitative comparison of the spatial-temporal predicted velocity field (Y-direction) and its error field for USFNO-FConvLSTM, Unet-FConvLSTM, and CNN-FConvLSTM models. The plots presented at dimensionless time instants $tU_0/D$=207, 216, 227 and 241 for two tandem-cylinders. The black dotted circle indicates the cylindrical location of the labeled field.

*4.4. The performance of USFNO-based hybrid neural network*

In Section 4.2 and Section 4.3, the specific network parameters and architecture of the novel USFNO-FConvLSTM model are determined by introducing the VIV of two tandem-cylinders as the FSI benchmark case. The modeling reliability and superiority are demonstrated in contrast to other models also proposed in this paper. On the basis of the above studies, two VIV benchmark projects, for instance, an isolated cylinder and two tandem-cylinders with 1DOF, are applied to analyze and verify the excellent predictive performance of the innovative USFNO-based hybrid neural network model (in Figure 1) to spatial-temporal evolution with high-nonlinear physical features. The single-cylinder case among them is to illustrate the robustness and generalization of the FSI neural network model for different engineering applications. The initial flow fields settings and oscillator properties for these selected cases are described in Section 3.2. It should be pointed out that in order to construct more appropriate FSI process and structural responses, we omit the time period when the initial vibration is not obvious, and set the re-selected initial vibration time to $tU_0/D$=0. This results in a difference from the time step range in Figure 11 and Section 4.3. Detailed research is carried out below.

*4.4.1. VIV modeling of an isolated cylinder*

For an isolated oscillator, 1400 consecutive time snapshots distributed in the range of $tU_0/D$ = [0, 280] with equal spacing are divided into training and predicting datasets successively in a ratio of 4:3. For training, the learning rate has the initial value of 0.001 and is dynamically adjusted by exponential decay policy with attenuation base of 0.977 based on 200 epochs. The time forecasting of the prediction region is derived by reference to the training model. The comparison of vibration displacements and corresponding velocities in cross-flow direction between the USFNO-based hybrid neural network model (denoted as Case1) and other comparable baseline models (denoted as Case2 and Case3) is exhibited within $tU_0/D$ = [0, 280] in Figure 21 and Figure 22, respectively. Of particular note is that the flow field modules of the chosen basic FSI models (Case2 and Case3) are both on the basis of CNN-FConvLSTM architecture (Figure A.1 in Appendix) and respectively belong to the multiple- and single-time step structure. We designate these two models as CNN-based FSI models. Among them, the neural framework with multiple-time steps is identical to the USFNO-based hybrid neural network model, differing only in the substitution of the CNN module for the USFNO. In addition, the information interaction between the fluid flow and structural responses of the single-time step model is consistent with the processing method in Han et al. (2022). That is, physical information is incorporated into the fluid network through additional channels. Figure 23 corresponds to the visual qualitative comparison of pressure fields including the labeled data, predicted results, and their relative errors, across five progressive time steps in the prediction phase for the above three FSI models. Table 3 summarizes the detailed errors in the flow field and structural responses during both training and predicting processes. While differences between USFNO and CNN were already compared in Section 4.3, this additional comparison further demonstrates the superiority of the USFNO-based hybrid neural model in VIV modeling when coupled with structural responses.

As shown in Figure 21 and Figure 22, the instantaneous vibration displacement and its velocity trained and predicted by the novel USFNO-based hybrid neural network model, has a consistent fitting trend with the labeled results of numerical simulation. Specifically, as $tU_0/D$ is increased from 160 to 280 in the prediction period, there is no apparent or even negligible phase difference between predicted and labeled curves, except the forecast deviation for a few amplitudes. Furthermore, an analogous phenomenon revealing the superior predictive performance of the USFNO-based hybrid network model can also be distinctly observed in Figure 23. That is, in spite of the forecast effect of the flow field has slight variations with the advance of the time sequences, the USFNO-based hybrid neural network model still maintains high prediction accuracy for complex physical fields, whether it is for the global flow field or the localized flow fields corresponding to the dynamic boundary of vibrating single-cylinder. This underscores its robustness in handling intricate FSI over extended periods. For instance, the average MRE during training is 0.51%, while for the prediction state, it stands at 5.30%. As a consequence, our proposed model has outstanding forecasting ability for both structural responses and flow field evolutions. This capability is even more prominent by comparing it with the other two CNN-based FSI models, i.e., Case2 and Case3. As is shown in Figure 21(a) and Figure 22(a), in the initial training, all three cases, especially Case3, generate an unsatisfactory imitative effect due to the vibrating instability and irregularity of the initial phase of VIV. Relatively, the predictive result of Case2 is acceptable. In light of the conclusions drawn from Figure 15 to Figure 20, owing to the inadequacies of the CNN architecture in capturing the dynamic evolution of sequential flow fields, it can be observed in Figure 21 - Figure 23 that even employing a comparable multi-time step strategy, the modeling performance of CNN-based FSI model remains inferior to that of the USFNO-based one. For instance, in the initial training (in Figure 21(b) and Figure 22(b)), with the increase of $tU_0/D$, a weak phase advance appears in Case2, which is contrary to the phase lag phenomenon in the last predicting state (in Figure 21(c) and Figure 22(c)). An evident local error is shown around the oscillator in the forecasting pressure field (6$^{th}$ column of Figure 23) for Case2. Even though the errors are localized, it is essential to evaluate the modeling reliability of the FSI model, particularly in predicting flow fields in the vicinity of dynamic boundaries. In comparing Case2 and Case3, the former shows more acceptable results, meaning that the multi-time step introduced in the proposed FSI neural model is beneficial to the representation of nonlinear physical fields. Besides that, Case3 with a single-time step presents the obvious phase deviation between its structural reflection and labeled data (in Figure 21(b,c)), meanwhile, displaying the unacceptable local and global pressure field error, as shown in the 7[th] column of Figure 23. Therefore, the proposed USFNO-based hybrid neural network model is more appropriate for FSI modeling for an isolated oscillator than other CNN-based FSI models proposed in this paper or modified on the basis of the existing models.

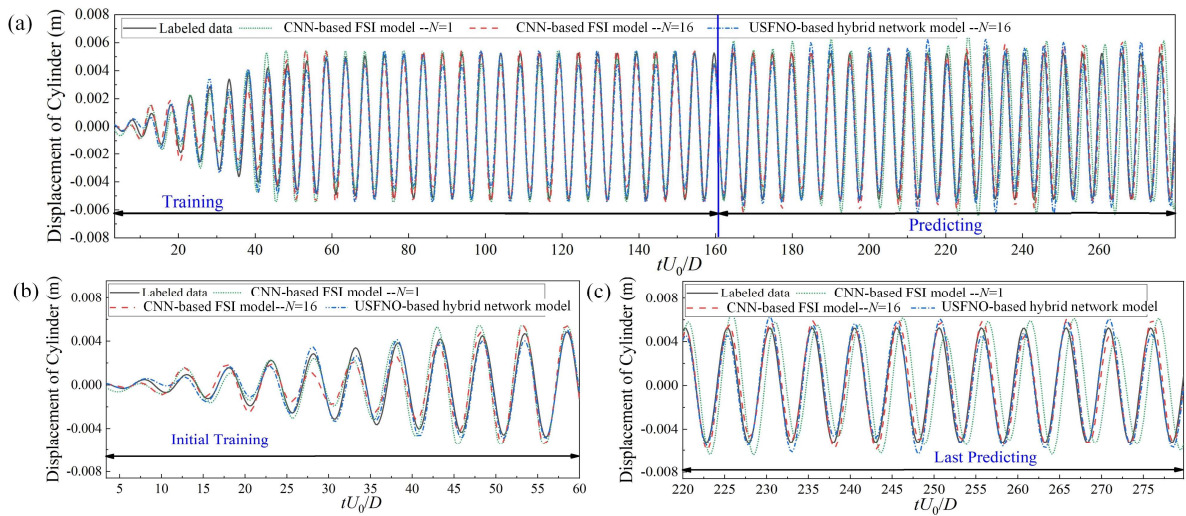

**Figure 21.** Comparison of displacement evolution trained and predicted by the CNN-based FSI model ($N$=1), CNN-based FSI model ($N$=16), and USFNO-based hybrid neural network model ($N$=16) for an isolated cylinder: (a) entire evolutionary process (b) initial training (c) last predicting.

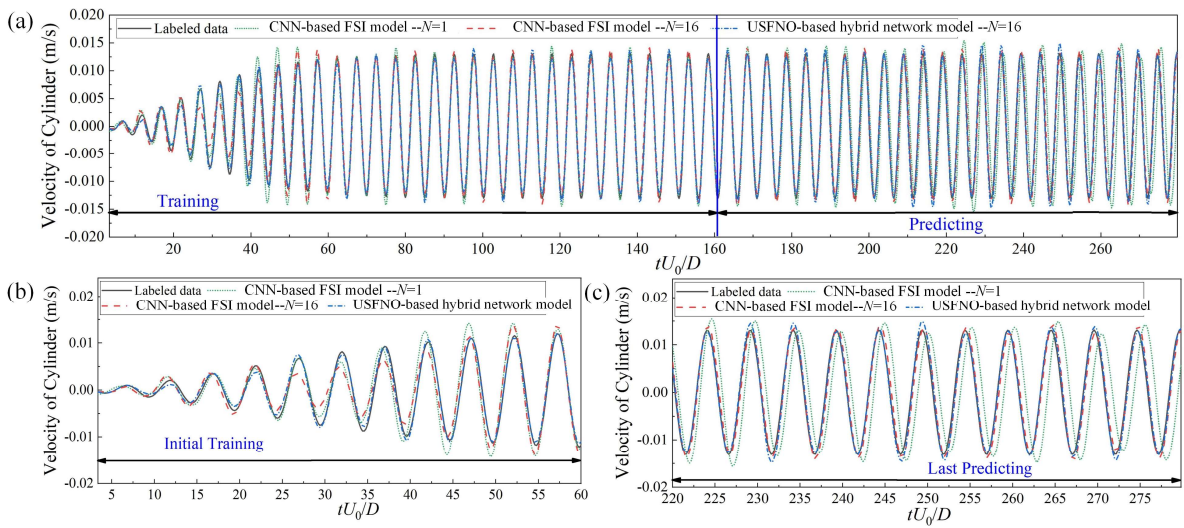

**Figure 22.** Comparison of velocity evolution trained and predicted by the CNN-based FSI model ($N$=1), CNN-based FSI model ($N$=16), and USFNO-based hybrid neural network model ($N$=16) for an isolated cylinder: (a) entire evolutionary process (b) initial training (c) last predicting.

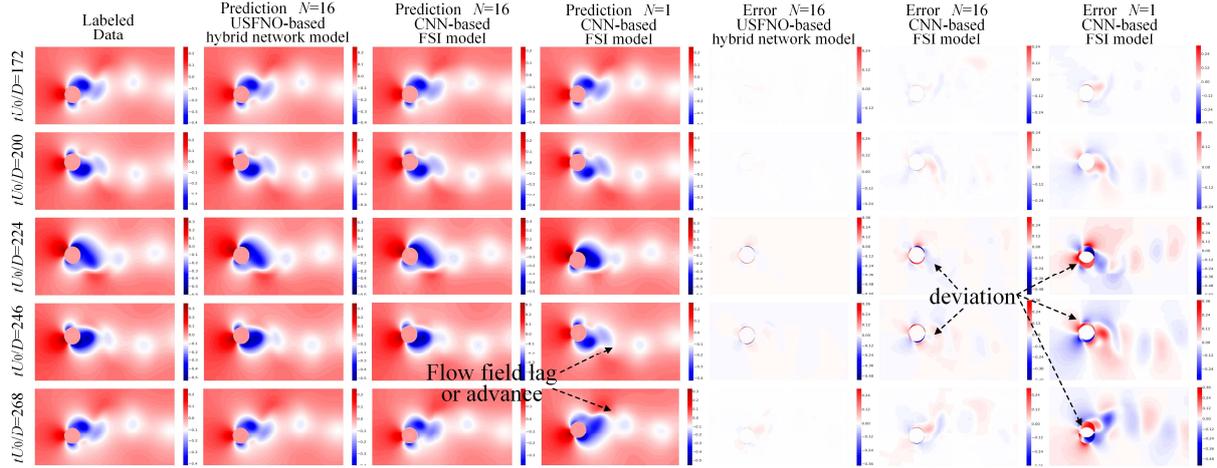

**Figure 23.** Qualitative comparison of the spatial-temporal predicted pressure field and its error field using the CNN-based FSI model ($N$=1), CNN-based FSI model ($N$=16), and USFNO-based hybrid neural network model ($N$=16) for an isolated cylinder. The plots presented at dimensionless time instants $tU_0/D$=172, 200, 224, 246 and 268.

**Table 3.** Training and predicting errors of flow field and structural responses for different models of an isolated cylinder.

| Process | Case | Pressure Field | | | Displacement [m] | | Velocity [m/s] | |
|---|---|---|---|---|---|---|---|---|
| | | MAE | MRE (%) | RMSE | MAE | RMSE | MAE | RMSE |
| Training | Case1 | 0.000697 | 0.51 | 0.0023 | 0.000407 | -- | 0.000582 | -- |
| | Case2 | 0.0130 | 8.98 | 0.0182 | 0.000513 | -- | 0.000992 | -- |
| | Case3 | 0.0085 | 5.92 | 0.0144 | 0.000566 | -- | 0.000991 | -- |
| Predicting | Case1 | 0.0084 | 5.30 | 0.0157 | 0.000498 | 0.000577 | 0.000706 | 0.000880 |
| | Case2 | 0.0180 | 11.02 | 0.0256 | 0.000571 | 0.000692 | 0.001111 | 0.001325 |
| | Case3 | 0.0680 | 41.23 | 0.0656 | 0.002069 | 0.002613 | 0.004980 | 0.006327 |

*4.4.2. VIV modeling of two cylinders in tandem*

For a multiple-oscillator system, 1200 consecutive time snapshots evenly distributed within $tU_0/D$ = [0, 240] are split into training and predicting datasets at a 2:1 ratio. For training, the initial learning rate is set as 0.0005. The exponential attenuation strategy with an attenuation base of 0.977 is used to dynamically adjust the subsequent learning rate. The time forecasting of the prediction region is derived by reference to the training model. Figure 24 displays the comparison of transient displacement evolutions predicted by the USFNO-based hybrid neural network model and simulated by numerical technology (labeled data) for both the upstream and downstream cylinders. Corresponding temporal analyses of vibration velocity, total lift force, and wall shear force are presented in Figure 25, Figure 27, and Figure 28, respectively. Figure 26 shows the spatial visualization of the forecasted pressure fields along with the absolute errors in comparison to the labeled values. To reflect the superiority of the USFNO-based hybrid neural network model, it contrasts with the CNN-based FSI model equipped with a single time step and FConvLSTM module.

In Figure 24, the instantaneous displacement derived by the USFNO-based hybrid neural network model displays a high precision fitting tendency with the labeled curve, covering but not limited to both the initial unstable vibration and the later periodic vibration phases. Corresponding quantitative evaluation indicators are summarized in Table 4 and Table 5. For instance, the MAE averaged over both the training and predicting phases for both the upstream and downstream oscillators is below 0.0006, which is an order of magnitude smaller than the maximum labeled displacement, proving the robust modeling capability of our proposed USFNO-based hybrid neural network model. We also ascribe this ascendant performance to integrating the gradient-based image processing (GIP) method in training, which contributes to rectifying the vibration displacement deviations during the iteration of FSI modeling. The GIP method computes the morphological gradient matrix of the flow field images, capturing gradient variations between each flow field point and its adjacent ones. A pronounced gradient can be observed at the cylindrical boundary, attributable to both the zero padding within the cylinder and the existence of alternating high- and low-pressure zones around the cylinder, a consequence of the vortex shedding phenomenon. In the latter stages of training, with highly accurate flow field predictions, we can determine the spatial position of oscillators utilizing the GIP technology. For training, as the epoch exceeds 180, the better one between the computationally derived vibration displacement and that captured via gradient analysis is selected to proceed with FSI modeling. In the predicting phase, the average of the aforementioned two forms is adopted as the temporal displacement for structural response. For the CNN-based FSI model with a single time step, the curve-fitting trend remains reasonable within the initial training ($0 < tU_0/D < 36$). Nevertheless, owing to instabilities in the initial flow field, the peak displacement of each vibrating period for upstream oscillator is attenuated within the interval $tU_0/D$ = [36, 45] (in Figure 24(a)). This nonlinear physical evolution induces substantial phase deviations in subsequent forecasts. Meanwhile, owing to the interaction between two tandem-cylinders, the time-accumulated errors from the upstream cylinder will indirectly affect the displacement prediction accuracy of the downstream cylinder, as shown in Figure 24(b). Simultaneously, the amplitude fluctuates periodically with the advance of the time series (i.e., amplitude envelope is in harmonic form), as presented in Figure 24. Moreover, for upstream cylinder, a vibrating divergence phenomenon is appeared since the prediction process due to the error accumulation (in Figure 24(a)). Therefore, the CNN-based FSI model is not applicable to construct the complicated displacement evolution of tandem cylinders.

Vibration velocity, as another indispensable evaluation indicator, is introduced to reflect the structural derivation ability of the FSI models. For the USFNO-based hybrid neural network model, in contrast to the excellent displacement prediction in Figure 24, velocity evolution exhibits a slightly inferior forecast because the instantaneous derivation speed cannot be stepwise corrected. More specifically, Figure 25 illustrates that the modeled oscillatory velocity shows an amplified amplitude compared to the labeled data. Nevertheless, there is no obvious phase deviation between the whole fitted curve and the labeled one, that is, the amplitude of each period corresponds to a similar $tU_0/D$. The above phenomena indicate that the coupled interactions between multiple cylinders induce a heightened degree of nonlinear physical complexity to the instantaneous flow field. Involving the USFNO-based hybrid neural network model, compared with the isolated cylinder application, VIV modeling for tandem cylinders shows slightly inferior predictive capability due to the increased dynamic boundaries augmenting the learnable features in the flow field model, thereby escalating complexity. For the CNN-based model (in Figure 25), the fitting trend develops from coincidence to phase difference as $tU_0/D$ increases, and the velocity divergence of the upstream cylinder also evolves in the late predicting stage, which is consistent with its displacement in Figure 24.

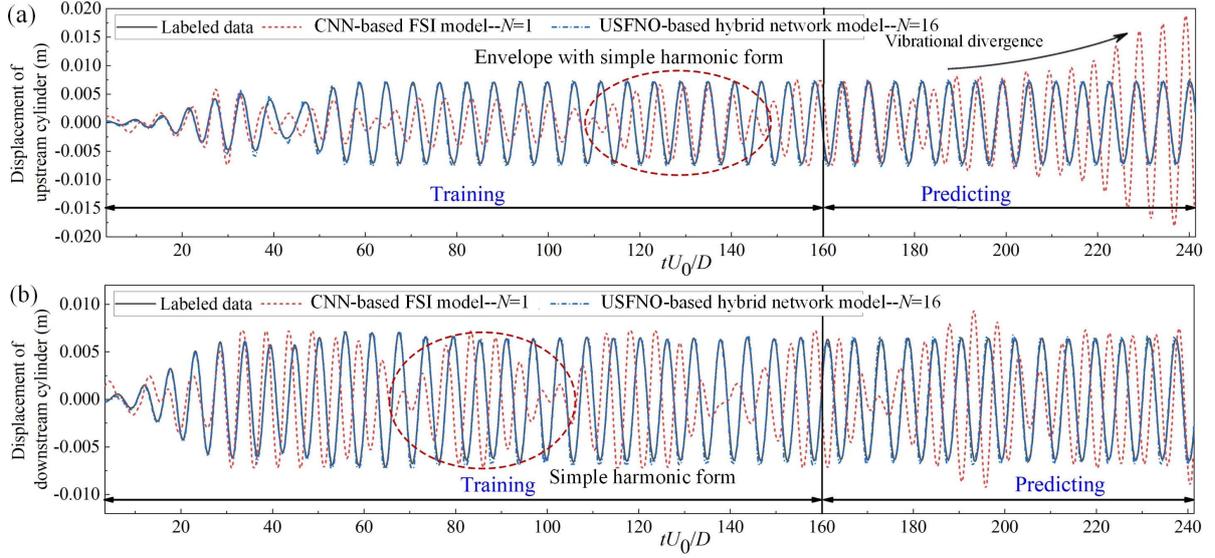

**Figure 24.** Transient vibration displacement of the CNN-based FSI model ($N=1$) and USFNO-based hybrid network model ($N=16$): (a) upstream cylinder (b) downstream cylinder.

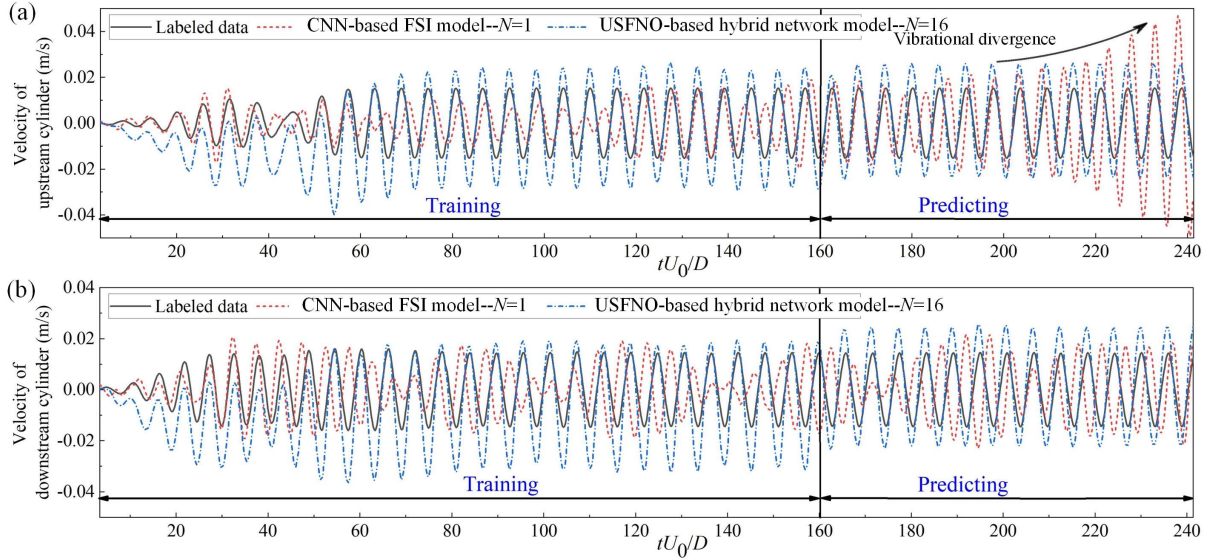

**Figure 25.** Transient vibration velocity of the CNN-based FSI model ($N=1$) and USFNO-based hybrid network model ($N=16$): (a) upstream cylinder (b) downstream cylinder.

In addition to the structural responses analyzed above, whether the flow field prediction is accurate or within the acceptable error range is of pivotal importance for evaluating the modeling performance of the proposed USFNO-based hybrid neural network model. For reasonable assessment, the instantaneous prediction fields and corresponding relative error distributions of five discontinuous time steps arranged by time sequence are selected in the range of $tU_0/D = [160, 240]$, as shown in Figure 26. It is evident that in the initial prediction, the pressure fields predicted by the USFNO-based hybrid neural network model (2$^{nd}$ column of Figure 26) are consistent with the labeled data (1$^{st}$ column of Figure 26). That is, almost all the flow features are captured with high fidelity, including cylindrical boundary domain and surrounding flow field. As the time sequence advances from $tU_0/D = 164$ to $tU_0/D = 232$, relative errors reveal more distinct, with a significant enhancement in the deviations observed in the vicinity of the cylindrical surface. However, the above modeling defects are acceptable and insignificant in contrast to the CNN-based FSI model. For instance, three quantitative evaluation indicators obtained by the USFNO-based hybrid neural network model in the whole prediction phase have the averages of MAE = 0.0159, MRE = 14.46%, and RMSE = 0.0209, respectively. The corresponding training phase has lower errors, as listed in Table 6. The above satisfactory phenomena indicate that the USFNO-based hybrid neural network model has a high degree of accuracy in predicting the evolution of the flow field surrounding the two oscillators, as well as capturing critical boundary information characterized by high gradients, all while accommodating the motion of the structure. On the contrary, the CNN-based FSI model presents a non-negligible defect in forecasting the pressure distribution around the downstream cylinder (3$^{rd}$ and 5$^{th}$ columns in Figure 26) during the initial stage (i.e., $tU_0/D$ = 164 and 184). With increasement of $tU_0/D$, this defect reflects more visibly, which is manifested in the inability to correctly capture the transient positions of the oscillators and the surrounding mutual pressure field. Quantificationally, the average MRE across the prediction range reaches as high as 45.72% (in Table 6), meaning that the CNN-based FSI model with a single time step is not suitable for modeling the flow field of two tandem cylinders.

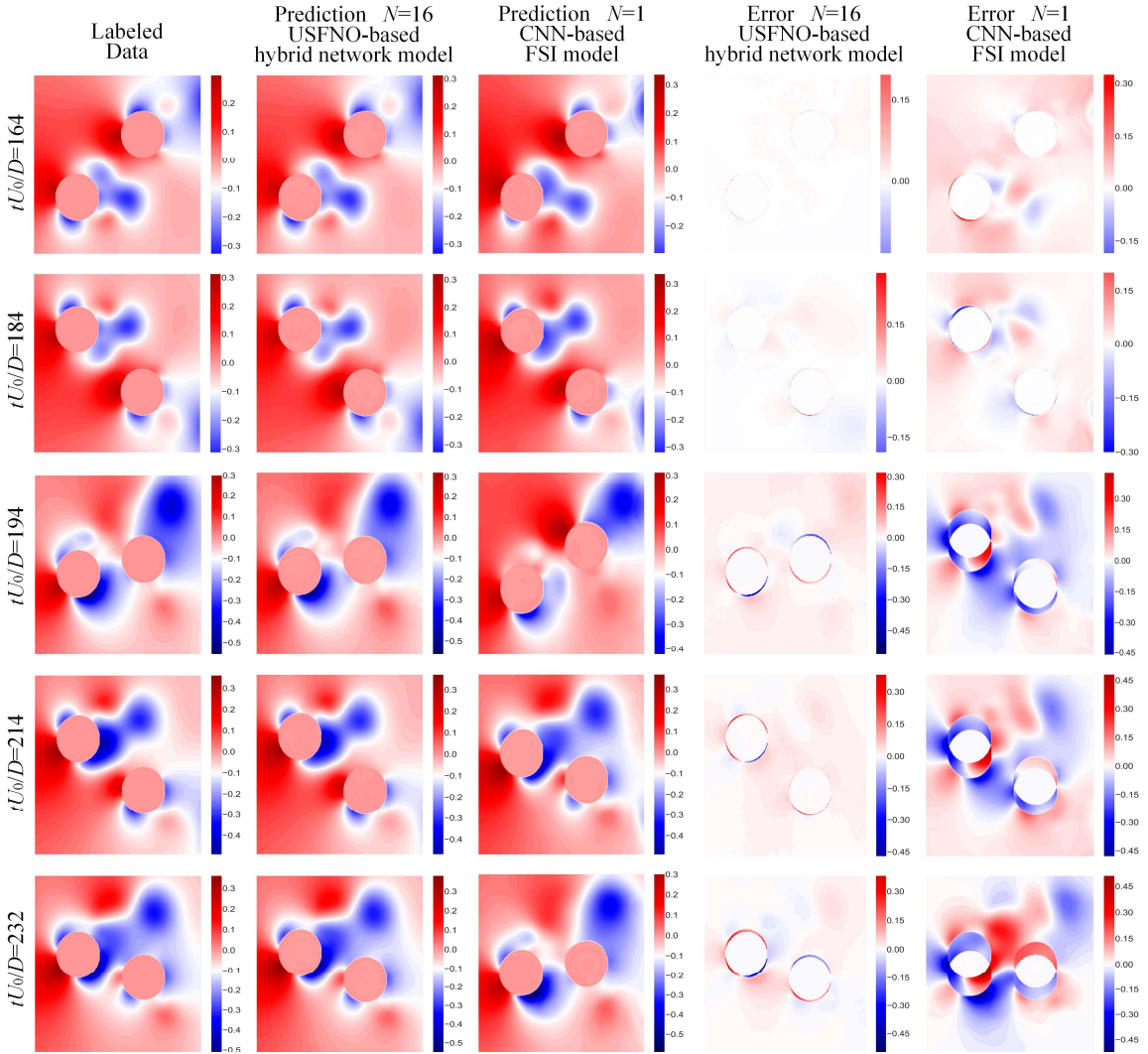

**Figure 26.** Qualitative comparison of the spatial-temporal predicted pressure field and its error field using the CNN-based FSI model (*N*=1) and USFNO-based hybrid neural network model (*N*=16) for two cylinders in tandem arrangement. The plots presented at dimensionless time instants $tU_0/D$=164, 184, 194, 214 and 232.

In Figure 27(a), for the upstream oscillator, apart from the initial training stage, the total lift derived from the transient pressure field and displacement predicting by the USFNO-based hybrid neural network model has a consistent continuous trend with the labeled data in terms of both evolutionary period and amplitude forces. The incomplete fitting is associated with the predictive precision of the pressure field and the deviation in derived displacement, which relates to the lift-solving principle illustrated in Figure 9. For the downstream oscillator, as drawn in Figure 27(b), the lift inferred by the USFNO-based hybrid neural network model also exhibits a similar variation tendency as the labeled one, but with a diminished fitting accuracy when juxtaposed against the observations made for the upstream cylinder. The relative weak fitting for downstream one is manifested as follows: (1) greater predictive lift amplitude deviation relative to the labeled results; (2) more persistent phase of initial aperiodic vibration; (3) phase deviation occurs in the late forecast period. However, these phenomena are reasonable and within the acceptable range. Due to the compact spacing of $L=1.5D$, the complex wake and vortex shedding induced by the upstream cylinder directly leads to the physic field with high nonlinearity for the downstream cylinder, making it more difficult to accurately predict the cylindrical boundary flow field, thereby diminishing the precision of lift derivation. Nevertheless, the integrating lift of the CNN-based FSI model, in comparison, is far less predictable, especially for downstream oscillator as expected. It deviates by half a vibration period and forms a series of lower force peaks, as shown by the red dashed line in Figure 27(b). Furthermore, the discrepancy in the total lift force prediction between the USFNO-based hybrid neural network model with multi-time steps and the CNN-based FSI model with a single-time step is obvious due to imprecise displacement and pressure field predictions. As shown in Figure 25 and Figure 27, whether the vibration velocity or the total lift force predicted by the USFNO-based hybrid neural network model, discrepancies from the labeled data during the initial training phase are visually manifested as a deviation towards the negative direction. This is owing to the instability of cylindrical oscillation and the intensity of flow field variation.

The time evolutions of wall shear forces calculated at discrete wall shear stress points predicted by the MLP model are compared in Figure 28. It can be observed that the applied MLP is beneficial to accurately predict the wall shear force, both for evolutionary trend and peak value. The only shortcoming is that the slight phase deviation appears between the prediction and label as $tU_0/D$ advances due to the error accumulation, but is within acceptable limits.

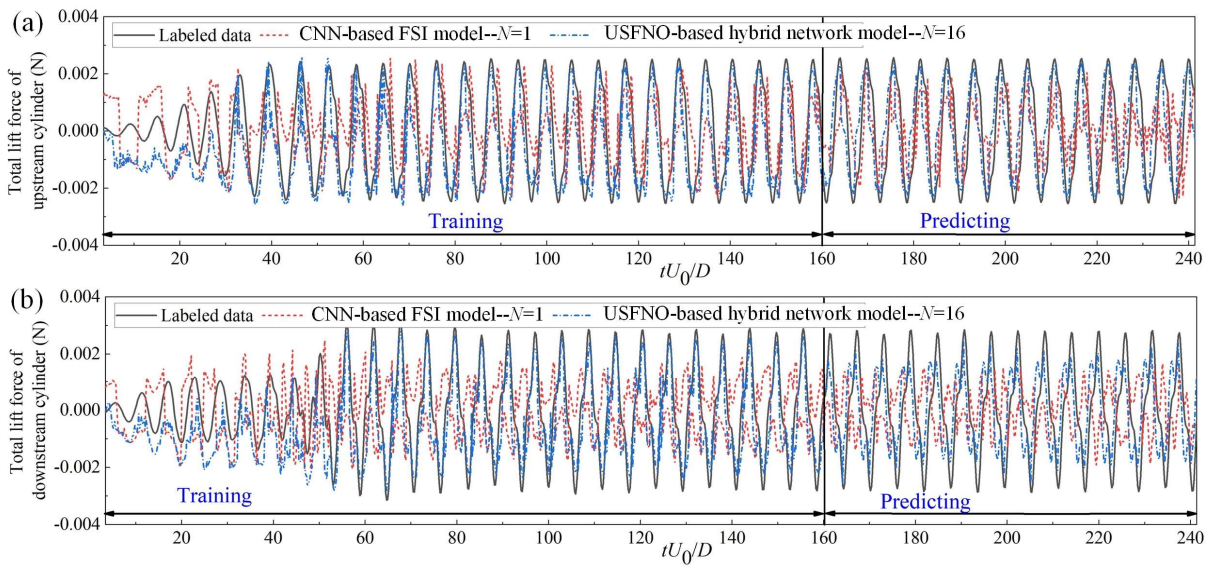

**Figure 27.** Transient lift force of the CNN-based FSI model ($N=1$) and USFNO-based hybrid network model ($N=16$): (a) upstream cylinder (b) downstream cylinder.

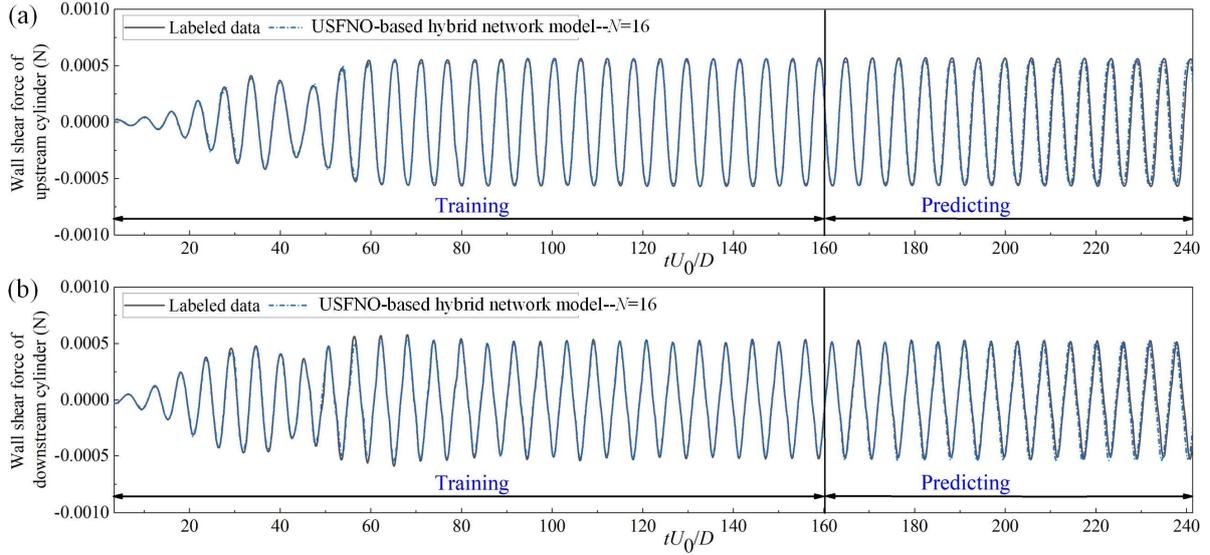

**Figure 28.** Wall shear force of the USFNO-based hybrid network model ($N$=16): (a) upstream cylinder (b) downstream cylinder.

To sum up, the proposed USFNO-based hybrid neural network model can satisfactorily complete the spatial-temporal prediction of the multi-cylinder system, i.e., flow field evolution and structural responses. Moreover, its modeling performance is far superior to the CNN-based FSI model under the same experimental conditions.

**Table 4.** Training and predicting errors of structural responses for different models of upstream cylinder.

| Process | Method | Displacement (m) | | Velocity (m/s) | | Lift Force (N) | |
|---|---|---|---|---|---|---|---|
| | | MAE | RMSE | MAE | RMSE | MAE | RMSE |
| Training | Case1 | 0.00030 | -- | 0.00848 | -- | 0.00087 | -- |
| | Case2 | 0.00343 | -- | 0.00764 | -- | 0.00127 | -- |
| Predicting | Case1 | 0.00055 | 0.00061 | 0.00665 | 0.00751 | 0.00099 | 0.00116 |
| | Case2 | 0.00634 | 0.00797 | 0.01591 | 0.01987 | 0.00116 | 0.00138 |

**Table 5.** Training and predicting errors of structural responses for different models of downstream cylinder.

| Process | Method | Displacement (m) | | Velocity (m/s) | | Lift Force (N) | |
|---|---|---|---|---|---|---|---|
| | | MAE | RMSE | MAE | RMSE | MAE | RMSE |
| Training | Case1 | 0.00028 | -- | 0.00902 | -- | 0.00084 | -- |
| | Case2 | 0.00520 | -- | 0.01155 | -- | 0.00176 | -- |
| Predicting | Case1 | 0.00051 | 0.00060 | 0.00706 | 0.00791 | 0.00112 | 0.00126 |
| | Case2 | 0.00657 | 0.00773 | 0.01466 | 0.01752 | 0.00180 | 0.00214 |

**Table 6.** Training and predicting errors of the flow field and wall shear force for different models of two cylinders in tandem.

| Process | Method | Pressure Field | | | Wall shear stress-Y1 direction | | | Wall shear stress-Y2 direction | | |
|---|---|---|---|---|---|---|---|---|---|---|
| | | MAE | MRE (%) | RMSE | MAE | MRE (%) | RMSE | MAE | MRE (%) | RMSE |
| Training | Case1 | 0.0012 | 1.11 | 0.0035 | 0.0007 | 3.28 | 0.0009 | 0.0011 | 7.31 | 0.0015 |
| Training | Case2 | 0.0088 | 8.32 | 0.0122 | -- | -- | -- | -- | -- | -- |
| Predicting | Case1 | 0.0159 | 14.46 | 0.0209 | 0.0046 | 21.01 | 0.0060 | 0.0040 | 26.47 | 0.0057 |
| Predicting | Case2 | 0.0532 | 45.72 | 0.0610 | -- | -- | -- | -- | -- | -- |

## 5. Conclusions

This paper investigated the combination of the Fourier neural operator and modified convolutional long-short term memory for spatial-temporal prediction of VIV of two cylinders in tandem arrangement and an isolated cylinder. We applied the CFD software combined with the overset mesh technique to obtain the accurate numerical simulation data of VIV used for modeling training and prediction. We proposed an innovative time series model for efficiently predicting flow field evolution problems with moving rigid bluff, and named it USFNO-FConvLSTM integrated by the FNO, FConvLSTM, and U-shaped framework. Furthermore, a well-designed FSI neural model, i.e., a USFNO-based neural network model consists of USFNO-based FSI recurrent units that contains fluid deep learning model and structural dynamics solver, was presented in this paper. By coupling the structural motion control equations and fluid model (composed of two separate data-driven solvers for flow field and wall shear stress subdomains) using force transfer and displacement fusion, an end-to-end, sequence-to-point training and predicting of FIS neural model for spatial-temporal modeling of VIV was conducted precisely and reliably.

The generated VIV data is verified to be accurate by comparing it with the existing results and is interpolated to adapt to network modeling. The effectiveness of the flow field features after interpolation is demonstrated through the proposed arc method. The accuracy and reliability of the USFNO-FConvLSTM model have been proven through the VIV of two cylinders in tandem arrangement. Appropriate sequence length ($N = 16$) and modified convolution LSTM model (FConvLSTM) are beneficial to improve the modeling accuracy of instantaneous physics fields. For the prediction phase, the minimum average MRE for pressure and velocity field are 9.56% and 3.3%. In comparison to the CNN- or Unet-based model with FConvLSTM and similar network parameters, the proposed USFNO-FConvLSTM indicates more superiority especially compared to the former.

The effectiveness and superiority of the USFNO-based hybrid neural network model have been demonstrated and exhibited through the VIV of two cylinders in tandem and a single cylinder with both 1DOF. The reasonable time sequence modeling of structural responses including vibration displacement and velocity, instantaneous flow field states, total fluid forces, and wall shear forces can be realized whether for training or prediction state. In contrast to the traditional CNN-based FSI model with a single time step, the USFNO-based hybrid neural network model with multi-time steps has higher capability for spatial-temporal prediction and accurate capture of dynamic boundary for multi-cylinder with complex nonlinear FSI features.

To sum up, the proposed FSI neural model has an excellent ability for constructing time sequence prediction of nonlinear complex dynamic structures. Future investigation can focus on the modeling generalization of multiple operating conditions, i.e., incoming flow velocity and spacing ratio between multi-cylinder. Furthermore, it has great potential to develop the FSI modeling for flexible structures with highly nonlinear deformation boundaries.

## Appendix A

The detailed architectures of the novel proposed CNN-FConvLSTM and Unet-FConvLSTM is shown in Figure A.1 and Figure A.2.

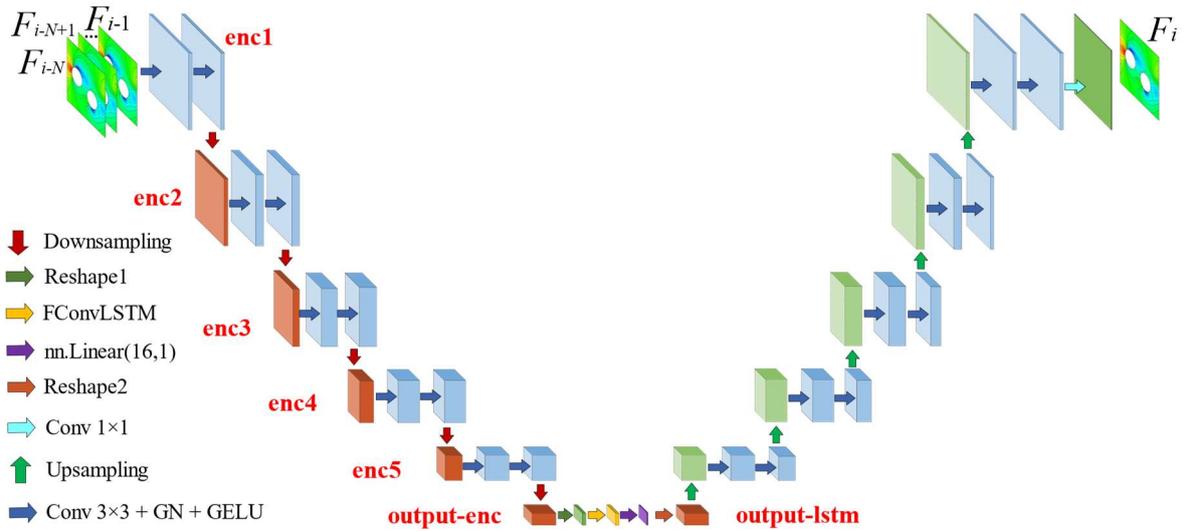

**Figure A.1.** The detailed architecture of the novel proposed CNN-FConvLSTM.

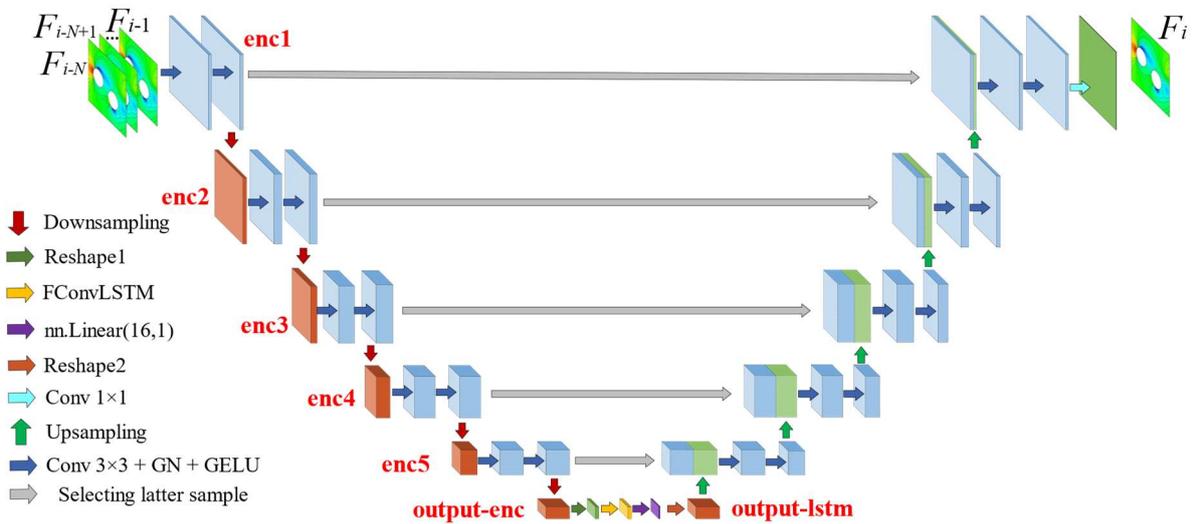

**Figure A.2.** The detailed architecture of the novel proposed Unet-FConvLSTM.

**Disclosure statement**

No potential conflict of interest was reported by the author(s).

**Funding**

This work was supported by the National Natural Science Foundation of China (No.92371206).

**CRediT authorship contribution statement**

Yanfang Lyu: Writing - original draft, Conceptualization, Data curation, Formal analysis, Investigation, Methodology, Software, Validation, Visualization. Yunyang Zhang: Writing - review and editing, Formal analysis, Investigation, Methodology. Zhiqiang Gong: Writing - review and editing, Investigation, Funding acquisition. Xiao Kang: Writing - review and editing, Supervision. Wen Yao: Writing - review and editing, Funding acquisition, Supervision. Yongmao Pei: Writing - review and editing, Supervision.

## Data availability statement

The data that support the findings of this study are available from the corresponding author upon reasonable request.